\documentstyle[preprint,aps,epsf]{revtex}
\begin{document}
\tighten

\draft
\preprint{ 
\parbox[t]{40mm}{{ADP-97-52/T278}\\
{UG-11/97}\\
}}
\title{Analytic structure of scalar composites
in the symmetric phase
of the gauged Nambu--Jona-Lasinio model
}
\author{V.P.~Gusynin$^{1,2}$ and M.~Reenders$^{1,3}$}
\address{$^1$
Bogolyubov Institute for Theoretical Physics,\\
252143 Kiev, Ukraine}
\address{$^2$
Centre for the Subatomic Structure of Matter,\\
University of Adelaide, South Australia 5005, Australia}
\address{$^3$
Institute for Theoretical Physics,\\
University of Groningen,
9747 AG Groningen, The Netherlands}

\date{December 1997}
\maketitle
\begin{abstract}
The gauged Nambu--Jona-Lasinio model in the quenched-ladder approximation
has nontrivial dynamics near a critical scaling region (critical curve)
separating a chiral symmetric and a dynamically chiral symmetry broken
phase. Scalar and pseudoscalar composites corresponding
to the four-fermion and gauge interactions become relevant degrees of 
freedom at long distances, which is reflected in the appearance of a large
anomalous dimension of the four-fermion operators. A method is introduced
for solving the Schwinger-Dyson equation for the Yukawa vertex
in specific kinematic regimes. This allows one to derive an analytic
expression for the scalar propagator, which is valid along the entire
critical curve. The mass and width of the scalar composites in the critical
scaling region are reexamined and the conformal phase transition
at gauge coupling $\alpha_0=\alpha_c$ is discussed.
\end{abstract}
\pacs{11.10.St, 11.15.Tk, 11.30.Qc, 11.30.Rd}
\section{Introduction}

The recent interest in the gauged Nambu--Jona-Lasinio (GNJL) model has been
stimulated by its importance for constructing extended technicolor
models (ETC) and the top-quark condensate model 
(for an introduction see Ref.~\cite{mirbook}). 

\begin{figure}[h]
\epsfxsize=10cm
\epsffile[120 320 370 600]{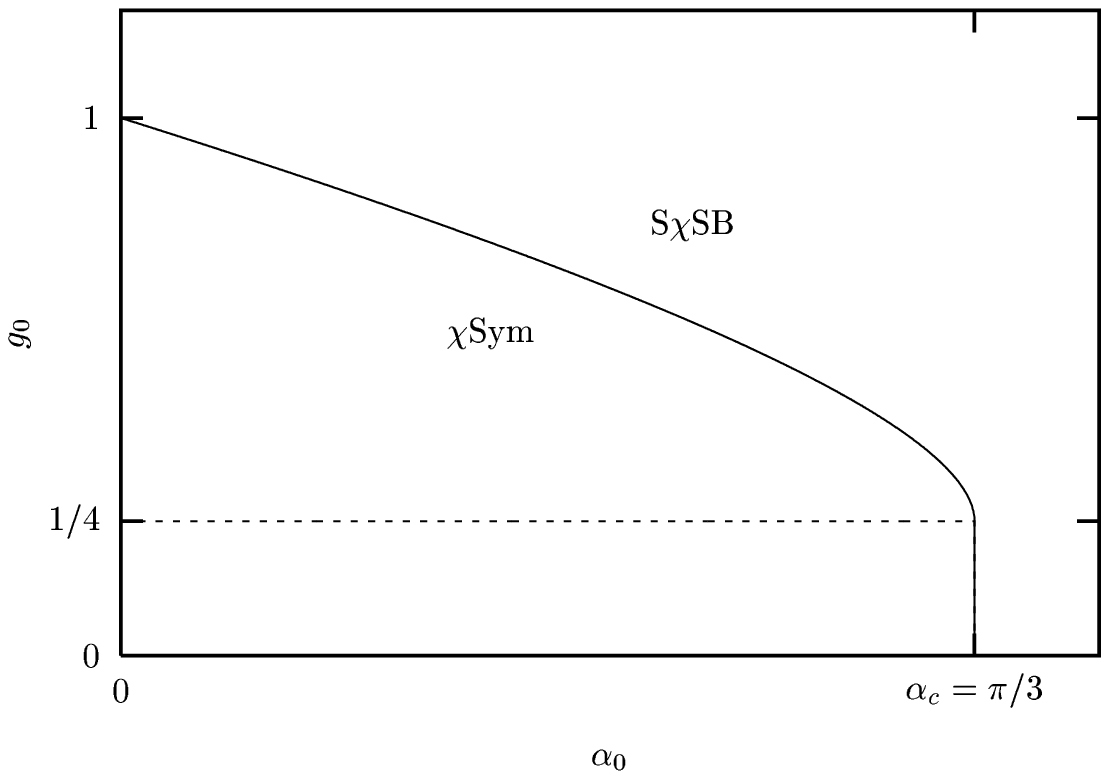}
\caption{The critical curve in the $(\alpha_0, g_0)$ plane.}
\label{critcurve}
\end{figure}

It is well known that in the quenched-ladder 
approximation\footnote{Also referred to as 
the quenched-planar approximation.} the GNJL model has a nontrivial 
phase structure \cite{komiya89} in the coupling constant plane 
$(g_0,\alpha_0)$, where $g_0\equiv 
G_0\Lambda^2/4\pi^2,\alpha_0\equiv e_0^2/4\pi$ as shown 
in Fig.~\ref{critcurve} ($\Lambda$ is a cutoff). The critical line is
\begin{equation}
g_c(\alpha_0)\equiv
\frac{1}{4}\left(1+\sqrt{1-\frac{\alpha_0}{\alpha_c}}\right)^2,\qquad
0\leq\alpha_0< \alpha_c=\frac{\pi}{3}
\label{critline1}
\end{equation}
at $g_0>{1\over4}$, and
\begin{equation}
 \alpha_0=\alpha_c
\label{critline2}
\end{equation}
at $g_0\leq{1\over4}$,
above which the gap equation for the fermion self-energy $\Sigma(p)$ has a 
nontrivial solution. Thus the chiral symmetry is dynamically broken, which 
implies the existence of a nonzero vacuum condensate 
$\langle\bar\psi\psi\rangle$. 
One end point $(g_0=1, \alpha_0=0)$ of the critical line corresponds 
to the ordinary NJL model, while the other one 
$(g_0=0,\alpha_0=\alpha_c=\pi/3)$ corresponds to pure QED. 
The interesting feature of the GNJL model is the
observation that naively irrelevant chiral invariant four-fermion
operators become relevant near the chiral phase transition \cite{balelo86}
due to the appearance of a large anomalous dimension $\gamma_m$ 
of the operator $\bar\psi \psi$
along the critical curve Eq.~(\ref{critline1}), 
$2\geq \gamma_m \geq 1$, and $\gamma_m=1$
along the part of the critical line with $\alpha_0=\alpha_c$ \cite{miya89}.
Therefore the GNJL model is believed to be
renormalizable as an interacting continuum theory near a critical scaling
region (critical curve) in the coupling plane separating a chiral symmetric
phase ($\chi S$) and a spontaneous chiral symmetry broken phase 
($S\chi SB$).

In the ladder approximation it has been shown 
\cite{gusmir,miransky,kotaya93}  that the GNJL model in four dimensions 
is indeed renormalizable, and that the anomalous dimension
\begin{eqnarray}
\gamma_m\equiv-\Lambda\frac{\partial \ln m_0}{\partial \Lambda}
\end{eqnarray}
is 
\begin{eqnarray}
\gamma_m&=&1-\omega+2\omega\frac{g_0}{g_c},
\qquad g_0<g_c={1\over4},\quad\omega=
\sqrt{1-\alpha_0/\alpha_c},\label{omega}\\
\gamma_m&=&1+\omega+2(g_0-g_c),\qquad g_0\geq g_c.
\label{anomdim}
\end{eqnarray}
Fine-tuning the coupling $g_0$ to $g_c$ in $S\chi SB$ phase 
in such a way that $m_d/\Lambda\ll 1$,
where $m_d\equiv\Sigma(0)$ is the dynamical mass of a fermion, a nontrivial
continuum limit ($m_d/\Lambda\rightarrow0$) can be reached just as in pure 
quenched QED
\cite{FGM,FGMS,MIR}. The spectrum of such a theory contains 
pseudoscalar ($\pi$) and scalar ($\sigma$) bound states which become light 
and dynamically active in the vicinity of the critical line. 
Since the phase transition is second order along the part 
Eq.~(\ref{critline1}) of the critical curve, scalar and pseudoscalar 
resonances have been shown to be produced on the symmetric 
side of the curve, whose masses approach
zero as the critical curve is approached \cite{aptewij91}. 
The part of the critical curve, Eq.~(\ref{critline2}), 
with $\alpha_0=\alpha_c$ is rather special. For example, an abrupt 
change of the spectrum of light excitations occurs when the line 
$\alpha_0=\alpha_c$, $g_0<1/4$ is crossed: while light scalar and 
pseudoscalar excitations still persist in the broken
phase, there are no such light excitations in the symmetric phase
\cite{miransky,ATW1}. 
A similar behavior has been revealed also in QED$_3$ \cite{ATW2}. 
This peculiar phase transition was referred to as a conformal phase 
transition (CPT) \cite{miya97}. 

\begin{figure}[h]
\epsfxsize=10cm
\epsffile[30 400 360 540]{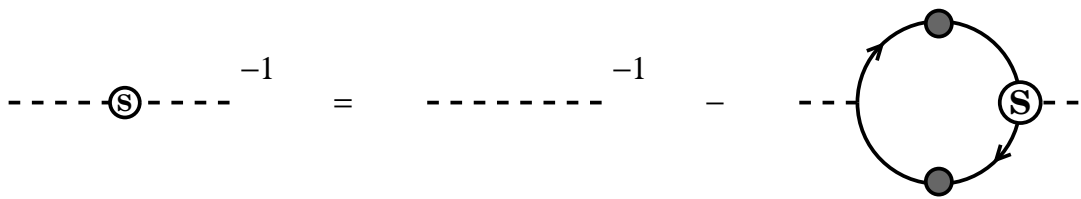}
\caption{The SDE for the scalar propagator $\Delta_{\rm S}(p)$.}
\label{fig_sde_scalar}
\end{figure}

\begin{figure}[h]
\epsfxsize=10cm
\epsffile[30 400 360 540]{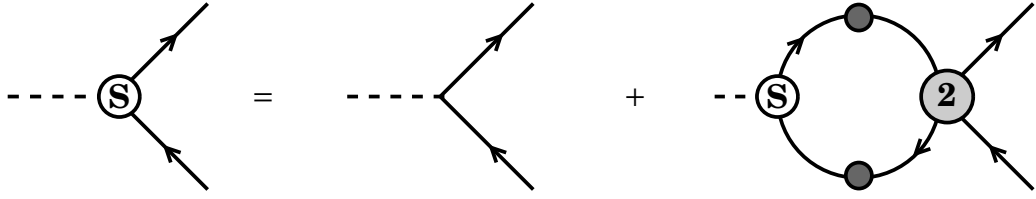}
\caption{The SDE for the scalar vertex ${\Gamma_{\rm S}}(p+q,p)$.
The shaded circle with the $2$ represents the two-fermion one-boson irreducible
fermion-fermion scattering kernel.}
\label{fig_sde_scalvert}
\end{figure}

\begin{figure}[h]
\epsfxsize=10cm
\epsffile[30 400 360 540]{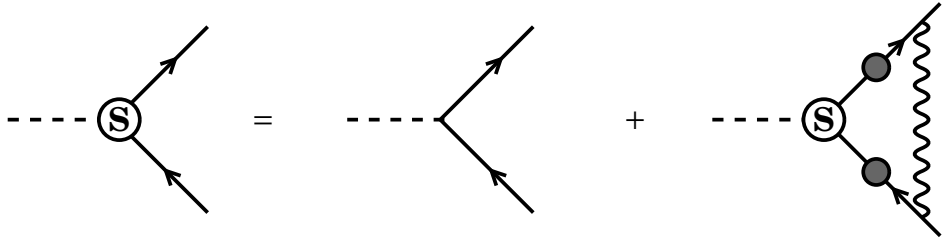}
\caption{The SDE for the scalar vertex in the quenched-ladder 
approximation.}
\label{fig_sde_scalvertladder}
\end{figure}

In this paper we study scalar composites ($\sigma$ and $\pi$ bosons)
in the symmetric phase of the GNJL model. Computing the scalar propagator 
(see Fig.~\ref{fig_sde_scalar}) requires knowledge of the full 
scalar-fermion-antifermion 
vertex ${\Gamma_{\rm S}}(p+q,p)$ which in turn satisfies the 
Bethe-Salpeter (BS) equation 
displayed in 
Fig.~\ref{fig_sde_scalvert}. We solve this BS equation in the ladder 
approximation (Fig.~\ref{fig_sde_scalvertladder}) but differ 
from the corresponding studies in 
Refs.~\cite{kotaya93,aptewij91} who used an approximation for 
${\Gamma_{\rm S}}(p+q,p)$ 
with zero boson momentum ($q=0$). 
A technique of expansion in the Chebyshev polynomials is introduced for 
solving the Yukawa vertex with nonzero boson momentum and consequently an 
explicit analytical expression is derived for the propagator of the 
$\sigma$ boson valid along the entire critical curve. 

Our main physical conclusions are the same as in 
Refs.~\cite{miransky,aptewij91,ATW1}: 
in the region $\alpha_0<\alpha_c$, $g_0>1/4$, 
in the symmetric phase, a spectrum of light resonances exists while 
at $\alpha_0<\alpha_c, g_0<1/4$ there are no light resonances. 
Having obtained an analytical expression for the scalar propagator, we can 
analytically continue it into the region $\alpha_0>\alpha_c$ and find 
light tachyons there, signaling the instability of the symmetric solution.

The plan of the present paper is as follows. After introducing the 
GNJL model (Sec.~\ref{sec_gnjl}) we solve the equation for the Yukawa 
vertex with nonzero boson momentum in Sec.~\ref{sec_scal_vertex} 
keeping only the zero order Chebyshev harmonics. 
In Sec.~\ref{scal_prop} we obtain an analytical expression for the 
boson propagator valid along 
the entire critical line and analyze its behavior in different 
asymptotical regimes. Sec.~\ref{sec_review} is devoted to comparing 
our results for the 
Yukawa vertex and boson propagator with the corresponding ones 
in Refs.~\cite{kotaya93,aptewij91}. 
In Sec.~\ref{res_near_crit} we discuss the behavior of the scalar 
propagator near the critical line (\ref{critline1}) in the symmetric phase 
and, in particular, the mass and the width of resonances. The analysis of 
the scalar composites near the critical line (\ref{critline2}) is given in 
Sec.~\ref{sec_cpt}, where we show 
the absence of light excitations at $\alpha_0\leq\alpha_c$ while 
analytically continuing the symmetric phase propagator into the region 
$\alpha_0\geq\alpha_c$ leads to the appearance of tachyonic states. 
We discuss this behavior from the viewpoint of the CPT conception 
proposed recently in Ref.~\cite{miya97}. We present our summary in 
Sec.~\ref{conclusion} and in Appendix~\ref{Chebyshev} we give an 
analysis of the contribution of higher order Chebyshev harmonics into 
the Yukawa vertex equation and scalar vacuum polarization. 

\section{The Gauged Nambu--Jona-Lasinio model}\label{sec_gnjl}

The gauged NJL model is described by the Lagrangian
\begin{equation}
{\cal L}_{\rm GNJL}=
-{1\over 4} F^2_{\mu\nu}+\bar \psi (i\gamma^{\mu}D_{\mu}-m_0)\psi
+{G_0\over2}[(\bar\psi \psi)^2+(\bar\psi i\gamma_5\psi)^2],
\label{La_1}
\end{equation}
where $D_{\mu}=\partial_{\mu}-ie_0A_{\mu}$ is a covariant derivative,
$e_0$ the gauge coupling constant,
and the last term is a chirally invariant four-fermion interaction
with $G_0$ the corresponding Fermi coupling constant. Another way to refer 
to this model is that it is QED with an additional 
four-fermion interaction. 
In the absence of a fermion mass term $m_0$ which breaks the chiral 
symmetry explicitly, the Lagrangian (\ref{La_1}) possesses a $U(1)$ 
gauge symmetry and a global $U_L(1)\times U_R(1)$
chiral symmetry. (It is not difficult  to extend our results for
$SU_L(N_f)\times SU_R(N_f)$ chiral symmetry with $N_f$ fermion flavors.)

Let us introduce the chiral fields $\sigma$ and $\pi$ rewriting the
Lagrangian (\ref{La_1}) in the form
\begin{equation}
{\cal L}_{\rm GNJL}=-{1\over 4}F^2_{\mu\nu}+\bar\psi i\gamma^\mu D_\mu\psi
-\bar\psi(\sigma+i\gamma_5\pi)\psi
-{1\over2G_0}(\sigma^2+\pi^2)+\frac{m_0}{G_0}\sigma.
\label{La_2}
\end{equation}
One can readily verify the equivalence of the Lagrangians (\ref{La_1})
and (\ref{La_2}) by just making use of the Euler-Lagrange equations. Further
we study mainly the chiral symmetric case with $m_0=0$.

With this Lagrangian we define the generating functional of the GNJL 
model by
\begin{eqnarray}
Z_{\rm GNJL}[J,\eta,\bar\eta,J_\sigma,J_\pi]=N\int
{\rm d}\mu\left(A,\bar\psi,\psi,\sigma,\pi\right)
\, \exp\left\{iS_{J} \right\},
\end{eqnarray}
where the ``measure'' is defined as
\begin{eqnarray}
{\rm d}\mu\left(A,\bar\psi,\psi,\sigma,\pi\right)\equiv
{\cal D}A{\cal D}\psi{\cal D}\bar\psi{\cal D}
\sigma{\cal D}\pi,
\end{eqnarray}
and where the source-dependent action $S_J$ is
\begin{eqnarray}
S_{J}=S_{\rm GNJL}+J_{\mu}A^{\mu}+\bar\eta\psi+\bar\psi\eta
+J_\sigma \sigma + J_\pi\pi+{\cal S}_{\Lambda}.
\end{eqnarray}
The last term ${\cal S}_{\Lambda}$ represents the ultraviolet
regulating part of the action.\footnote{ 
In practice we use a regularization by means of a hard ultraviolet cutoff
in the Euclidean momentum space integrals.
Moreover the axial anomaly is neglected.} 
The starting point to derive the Schwinger-Dyson equations (SDEs)
is the following formal identity:
\begin{eqnarray}
&&\int{\rm d}\mu\left(A,\bar\psi,\psi,\sigma,\pi\right)\,
\frac{\delta iS_{\rm J}}{\delta \phi(x)}
\exp\left\{iS_{J}\right\}=0,\quad \mbox{where}\quad
\phi=A,\psi,\bar\psi,\sigma,\pi,
\label{basicSDE}
\end{eqnarray}
from which the SDE for the propagators and vertices in momentum space 
can be obtained. For our purpose it is convenient to recall the SDEs for 
the scalar propagator and scalar vertex. The SDE equation for the scalar 
propagator is given by
\begin{eqnarray}
\Delta_{\rm S}^{-1}(p)=-\frac{1}{G_0}+\Pi_{\rm S}(p^2),
\label{delsinv}
\end{eqnarray}
where  the scalar vacuum polarization is
\begin{eqnarray}
\Pi_{\rm S}(p^2)=
i\int_{\Lambda}\frac{{\rm d}^4k}{(2\pi)^4}\,{\rm Tr}
\left[S(k+p){\Gamma_{\rm S}}(k+p,k)S(k) \right]
\label{sde_scalvac}
\end{eqnarray}
(see Fig.~\ref{fig_sde_scalar}), $S(k)$ is the full fermion propagator, and
${\Gamma_{\rm S}}(k+p,k)$ is the scalar fermion-antifermion vertex.
The absence of kinetic terms for the $\sigma$ and $\pi$ fields in the
Lagrangian is reflected in the constant bare propagator $-G_0$.

The scalar propagator is defined as the Fourier transform of the connected
part of the  correlator
\begin{eqnarray}
\langle0|T\left(\sigma(x)\sigma(y)\right)|0\rangle_{connected}=
\int\frac{{\rm d}^4k}{(2\pi)^4}{\,\rm e}^{-ik(x-y)}i\Delta_{\rm S}(k).
\end{eqnarray}
and the scalar vertex or Yukawa vertex ${\Gamma_{\rm S}}$ is defined as 
the Fourier transform of the amputated vertex of the three-point correlator
\begin{eqnarray}
\langle0|T
\left(\psi_{a}(x)\bar\psi_{b}(y) \sigma(z)\right) |0\rangle_{connected}
&=&\int\frac{{\rm d}^4k{\rm d}^4p}{(2\pi)^8}
{\,\rm e}^{-ik(x-z)+ip(y-z)}\nonumber\\
&\times&
\left[iS(k)(-i){\Gamma_{\rm S}}(k,p) iS(p)\right]_{ab}i\Delta_{\rm S}(k-p).
\end{eqnarray}
The full untruncated SDE for the scalar vertex in momentum space reads
\begin{eqnarray}
-i{\Gamma_{\rm S}}_{ab}(p+q,p)=
({-i\bf 1})_{ab}+&&\int\frac{{\rm d}^4r}{(2\pi)^4}
\left[iS(r+q) (-i){\Gamma_{\rm S}}(r+q,r) iS(r) \right]_{dc} \nonumber\\
&&\qquad\times(-ie_0^2)K_{cd,ab}(r,r+q,p+q)
\end{eqnarray}
(see Fig.~\ref{fig_sde_scalvert}), and the Bethe-Salpeter kernel $K$
is defined as the two-fermion one-boson irreducible fermion-fermion scattering
kernel. In the symmetric phase of the GNJL the pseudoscalar and scalar 
propagators are degenerate, so are the pseudoscalar vertex and 
scalar vertex.

\section{ Scalar Vertex in Quenched Ladder Approximation}
\label{sec_scal_vertex}

In this section we discuss the SDEs for the scalar propagator
and the scalar vertex in the well-known quenched-ladder approximation and 
introduce an approximation scheme for solving the SDE for the scalar 
vertex.
The ladder approximation is obtained by replacing the Bethe-Salpeter
kernel $K$ by the one photon exchange graph.
Furthermore the photon propagator is considered as quenched, i.e.,
vacuum polarization effects are turned off, and thus the gauge
coupling does not run.
In principle the Bethe-Salpeter kernel also contains scalar and 
pseudoscalar exchanges. One question is whether such exchanges can be 
neglected.
It is beyond the scope of this paper to give a complete answer 
to such questions, since the answer not only depends on the 
short-distance behavior of the full scalar propagators and Yukawa vertex
which is yet unsolved, but also on the representation of the chiral 
symmetry.  
In this respect, it is interesting to note that if
one includes the ladder-like one-scalar and one-pseudoscalar exchanges in the
truncation of the BS kernel $K$, and considers a chiral symmetry 
representation in which the number of scalars equals the number of 
pseudoscalars (thus both scalars and pseudoscalars in adjoint 
representation), then such contributions cancel each other exactly in 
the symmetric phase. Hence, provided scalars and pseudoscalars are 
considered  both in the adjoint representation of the chiral symmetry, 
the neglect of scalar and pseudoscalar 
exchanges in the kernel $K$ seems reasonable for the SDE 
for the Yukawa vertices ${\Gamma_{\rm S}}$ and ${\Gamma_{\rm P}}$.

The SDE equation for the scalar vertex in the ladder approximation 
can be written as
\begin{eqnarray}
{\Gamma_{\rm S}}(p+q,p)&=&{\bf 1}+ie_0^2\int\frac{{\rm d}^4r}{(2\pi)^4}
\gamma^\lambda S(r+q) 
{\Gamma_{\rm S}}(r+q,r) S(r) 
\gamma^\sigma D_{\lambda\sigma}(r-p)
\label{sde_ladder}
\end{eqnarray}
(see Fig.~\ref{fig_sde_scalvertladder}).
The SDE for the scalar propagator, Eq.~(\ref{sde_scalvac}), is left 
unchanged.
In the symmetric phase, the equation for the scalar vertex,
Eq.~(\ref{sde_ladder}), is a self-contained equation, if we note that 
in the Landau gauge the fermion propagator is $S(p)=1/\hat p$.
The SDEs in ladder approximation for scalar propagator
and vertex, Eqs.~(\ref{sde_scalvac}) and (\ref{sde_ladder}), 
have been studied extensively in the literature 
\cite{kotaya93,aptewij91,gukumi89,balelo89,balo92}, 
but mainly for the case of zero transfer boson momentum.

In what follows, we present a method for solving the scalar vertex with 
nonzero boson momentum. The starting point is a general structure of 
the scalar vertex and pseudoscalar vertex. The scalar and pseudoscalar 
vertices in momentum space can be decomposed over four spinor structures 
with dimensionless scalar functions in the following way:
\begin{eqnarray}
{\Gamma_{\rm S}}(k+q,k)&=&F^{({\rm s})}_1(k+q,k)+
\left(\hat q\hat k
-\hat k\hat q\right)
F^{({\rm s})}_2(k+q,k)\nonumber\\
&+& (\hat k+\hat q) F^{({\rm s})}_3(k+q,k)
+\hat k F^{({\rm s})}_4(k+q,k),\label{vertfiesdef}\\
{\Gamma_{\rm P}}(k+q,k)&=&(i\gamma_{5})\bigg[F^{({\rm p})}_1(k+q,k)+
\left(\hat q\hat k-\hat k\hat q\right)F^{({\rm p})}_2(k+q,k)\nonumber\\
&+&(\hat k+\hat q)F^{({\rm p})}_3(k+q,k)
+\hat k F^{({\rm p})}_4(k+q,k)\bigg],
\end{eqnarray}
where the scalar functions $F^{({\rm s})}_i$ and $F^{({\rm p})}_i$ 
depend on the squares of
the Minkowski momenta, $(k+q)^2$, $k^2$, $q^2$. Thus by use of 
the notation $F(k+q,k)$ we actually refer to a momentum 
dependence $F((k+q)^2,k^2,q^2)$.
Unless mentioned otherwise, our investigations will be focussed mainly
on the symmetric phase of the GNJL model.
In the symmetric phase, the equations for the scalar functions $F_3$
and $F_4$ decouple from the equations for $F_1$ and $F_2$. Moreover, 
$F_3$ and $F_4$ do not contribute in scalar and pseudoscalar vacuum 
polarizations. In fact the integral equations for these functions 
are homogeneous ones and in the symmetric phase we can always take the 
solution $F_3=F_4=0$ which is a consistent one. Furthermore the scalar 
and pseudoscalar vertex functions coincide,
i.e., $F^{(\rm s)}_i=F^{(\rm p)}_i$, $i=1,2$.
The consequence is that the scalar and pseudoscalar propagators are 
identical in the symmetric phase.
So the problem is reduced to solving a coupled set of integral equations
for two scalar functions $F_1$, $F_2$, which has the form 
(after making a standard Wick rotation \cite{tony})
\begin{eqnarray}
F_i(p+q,p)=\delta_{i1}
+\lambda \sum_{j=1}^2 \int_{\Lambda} {\rm d}r^2\,
\int\frac{{\rm d}\Omega_r}{2\pi^2}\,
K_{ij}(p,q,r) F_j(r+q,r),\qquad i=1,2, \label{sde_vertfies}
\end{eqnarray}
where $\lambda=3\alpha_0/4\pi$, and
\begin{eqnarray}
K_{11}(p,q,r)&=&
\frac{(r^2+q\cdot r)}{(r+q)^2 (r-p)^2},\label{k11}\\
K_{12}(p,q,r)&=&
\frac{2[(q\cdot r)^2-r^2 q^2]}{(r+q)^2 (r-p)^2},\label{k12}\\
K_{21}(p,q,r)&=&\frac{1}{6}
\frac{\kappa(p,q,r)}{(r+q)^2(r-p)^4[(p\cdot q)^2-p^2 q^2]},
\label{k21}\\
K_{22}(p,q,r)&=&\frac{1}{3}
\frac{\kappa(p,q,r)(r^2+q\cdot r)}{(r+q)^2(r-p)^4[(p\cdot q)^2-p^2 q^2]},
\label{k22}
\end{eqnarray}
with
\begin{eqnarray}
\kappa(p,q,r)&=&
p^2 p\cdot r q^2-p^2 p\cdot q q\cdot r-2 p\cdot q p\cdot r q\cdot r
+2 p^2 (q\cdot r)^2\nonumber\\
&+&2(p\cdot q)^2 r^2-2 p^2 q^2 r^2+p\cdot r q^2 r^2-p\cdot q q\cdot r r^2,
\end{eqnarray}
and $\int {\rm d}\Omega_r$ denotes the usual angular part of
the four-dimensional integration.

The equations Eq.~(\ref{sde_vertfies}) are still very complicated due to
the fact that the angular part of the integration cannot be performed
in explicit form, since the angular dependence of the Yukawa vertex 
is unknown. Without any further approximations it seems impossible to 
solve the equations analytically. Our primary interest is the scalar 
propagator defined by the vacuum polarization Eq.~(\ref{sde_scalvac}).
The equation for the scalar vacuum polarization is
\begin{eqnarray}
\Pi_{\rm S}(q^2)=\frac{1}{4\pi^2}
\int\limits_0^{\Lambda^2}
{\rm d} k^2\,\int\frac{{\rm d}\Omega_k}{2\pi^2}
\biggr[ A_1(k,q)  F_1(k+q,k)
+ A_2(k,q)F_2(k+q,k)\biggr], 
\label{sde_vacpol2}
\end{eqnarray}
where
\begin{eqnarray}
A_1(k,q)\equiv \frac{k^2+k\cdot q}{(k+q)^2},\qquad
A_2(k,q)\equiv\frac{2[(k\cdot q)^2-k^2 q^2]}{(k+q)^2}. \label{A1A2def}
\end{eqnarray}
The method to tackle the angular dependence is to expand in terms of 
Chebyshev polynomials of the second kind $U_n(x)$, a method which was 
used before, for instance, in Refs.~\cite{FGMS} and \cite{mawa96}.

We define the following expansions:
\begin{eqnarray}
F_1(p+q,p)&=&\sum_{n=0}^\infty f_n(p^2,q^2) U_n(\cos\alpha),\qquad
F_2(p+q,p)=\sum_{n=0}^\infty g_n(p^2,q^2) U_n(\cos\alpha),\\
A_1(p,q)&=&\sum_{n=0}^\infty a_n(p^2,q^2) U_n(\cos\alpha),\qquad
A_2(p,q)=\sum_{n=0}^\infty b_n(p^2,q^2) U_n(\cos\alpha),
\end{eqnarray}
and for the kernels, Eqs.~(\ref{k11})--(\ref{k22})
\begin{eqnarray}
K(p,q,r)&=&\sum_{n,m,l=0}^\infty K_{nml}(p^2,q^2,r^2) U_n(\cos\alpha)
U_m(\cos\beta)U_l(\cos\gamma),
\end{eqnarray}
where
\begin{eqnarray}
\cos \alpha=\frac{p\cdot q}{p q},\quad
\cos \beta=\frac{p\cdot r}{p r},\quad
\cos \gamma=\frac{q\cdot r}{q r}
\end{eqnarray}
(for the coefficients $K_{nml}(p^2,q^2,r^2)$ see Appendix~\ref{Chebyshev}).
After that the angular integration can be done explicitly leading to an 
infinite chain of equations for harmonics $f_n(p^2,q^2)$ and $g_n(p^2,q^2)$. 
The important thing is that only the harmonics $f_0$ contains an 
inhomogeneous term in the equation for it (the constant $1$ in 
Eq.~(\ref{sde_vertfies})), while other harmonics can be found iteratively 
once $f_0$ is computed. In other words, for the vertex function $F_1$ the 
scale is set by the bare vertex, i.e., such a function has nonhomogeneous 
ultraviolet boundary conditions. For the vertex function $F_2$ there is 
no such inhomogeneous term other than given indirectly by the coupling to 
vertex function $F_1$.

We assume that the scalar vertex function $F_1(p+q,p)$ depends only weakly 
on the angle between fermion and scalar-boson momentum $p\cdot q$,
so that an infinite set of equations for $f_n$ and $g_n$ 
is replaced by the equation for the zeroth-order Chebyshev 
coefficient function $f_0$ which we shall solve exactly. 
The main approximation is to replace
the Yukawa vertex by the angular average of vertex function $F_1$,
\begin{eqnarray}
{\Gamma_{\rm S}}(p+q,p)\approx {\bf 1} 
\int\frac{{\rm d}\Omega_p}{2\pi^2} F_1(p+q,p)
={\bf 1} f_0(p^2,q^2),
\label{canonic}
\end{eqnarray}
since Chebyshev polynomials of the second kind are precisely 
orthogonal with respect to such integration.
Then we write
\begin{eqnarray}
f_0(p^2,q^2)\equiv F_{\rm IR}(p^2,q^2)\theta(q^2-p^2)
+ F_{\rm UV}(p^2,q^2)\theta(p^2-q^2).\label{channelapprox}
\end{eqnarray}
The functions $F_{\rm IR}$ 
and $F_{\rm UV}$ are respectively referred to as the IR channel
(infrared), and the UV channel (ultraviolet).

If the scalar vertex indeed weakly depends on angle between
scalar-boson and fermion momentum $p$ flowing through 
the Yukawa vertex, these channel functions should have the limits
\begin{eqnarray}
\lim_{p^2\gg q^2} {\Gamma_{\rm S}}(p+q,p)&=&
{\bf 1}\lim_{p^2\gg q^2} F_{\rm UV}(p^2,q^2),\\
\lim_{q^2\gg p^2} {\Gamma_{\rm S}}(p+q,p)&=&
{\bf 1}\lim_{q^2\gg p^2} F_{\rm IR}(p^2,q^2),
\end{eqnarray}
i.e., the asymptotics of the scalar vertex are independent of the angle 
between $p$ and $q$. Hence the UV channel contains a limit of the Yukawa 
vertex with the boson momentum $q$ that is much less than both fermion momenta 
$(q\ll p)$, and the IR channel contains a limit of the vertex 
with the fermion momentum $p$ 
that is much less than the boson momentum $(q\gg p)$. The connection between 
the Yukawa vertex ${\Gamma_{\rm S}}$ and these two channel 
functions is illustrated in Fig.~\ref{two_channel_fig}.

The expansion in Chebyshev polynomials is discussed in detail 
in Appendix~\ref{Chebyshev}. Moreover, the error, due to our approximation, 
Eq.~(\ref{canonic}), in the computation of the scalar vacuum polarization, 
is estimated in that appendix. 

The zeroth-order Chebyshev or the two channel approximation 
of Eq.~(\ref{channelapprox}) gives the following
equation for the vertex function $f_0(p^2,q^2)$:
\begin{eqnarray}
f_0(p^2,q^2)=1+\lambda
\int\limits_0^{\Lambda^2} {\rm d}r^2\,N_0(r^2,p^2)
a_0(r^2,q^2) f_{0}(r^2,q^2),
\end{eqnarray}
where
\begin{eqnarray}
N_0(r^2,p^2)=\frac{\theta(r^2-p^2)}{r^2}+\frac{\theta(p^2-r^2)}{p^2}
\label{N0secIII}
\end{eqnarray}
and
\begin{eqnarray}
a_0(r^2,q^2)&=&\frac{1}{2}\left[\left(2-\frac{q^2}{r^2}\right)\theta(r^2-q^2)
+\frac{r^2}{q^2} \theta(q^2-r^2)\right]
\label{a0secIII}
\end{eqnarray}
(see also Appendix~\ref{Chebyshev}).
The equation for the scalar vacuum polarization 
in this approximation Eq.~(\ref{sde_vacpol2}) takes the form
\begin{eqnarray}
\Pi_{\rm S}(q^2)=
\frac{1}{4\pi^2}\int\limits_0^{\Lambda^2}
{\rm d} k^2\, a_0(k^2,q^2) f_0(k^2,q^2 ).\label{vacpol3}
\end{eqnarray}

With the Eqs.~(\ref{N0secIII}) and (\ref{a0secIII}) for $N_0$ and $a_0$,  
respectively, and the definition of the channel functions 
Eq.~(\ref{channelapprox}), we get two coupled integral equations for 
the IR-UV channels:
\begin{eqnarray}
F_{\rm IR}(s,t)&=_{_{_{_{\hspace{-5mm}{{ (s<t)}}}}}}&
1+\lambda\int\limits_0^s{\rm d}u\,\frac{u}{2st} F_{\rm IR}(u,t)
+\lambda\int\limits_s^t{\rm d}u\,\frac{1}{2t} F_{\rm IR}(u,t)
+\lambda\int\limits_t^{\Lambda^2}{\rm d}u\, 
\frac{2u-t}{2u^2} F_{\rm UV}(u,t),\label{fireq}\\
F_{\rm UV}(s,t)&=_{_{_{_{\hspace{-5mm}{{ (s>t)}}}}}}&
1+\lambda\int\limits_0^t{\rm d}u\,\frac{u}{2st} F_{\rm IR}(u,t)
+\lambda\int\limits_t^s{\rm d}u\,
\frac{2u-t}{2su}F_{\rm UV}(u,t)\nonumber\\
&+&\lambda\int\limits_s^{\Lambda^2}{\rm d}u\,
\frac{2u-t}{2u^2}F_{\rm UV}(u,t),
\label{fuveq}
\end{eqnarray} 
where $s=p^2$, $t=q^2$, and $u=r^2$. For the vacuum polarization, 
Eq.~(\ref{vacpol3}), we obtain the equation
\begin{eqnarray}
\Pi_{\rm S}(t)=
\frac{1}{4\pi^2}
\left[\int\limits_0^t {\rm d}u\,\frac{u}{2t} F_{\rm IR}(u,t)
+\int\limits_t^{\Lambda^2} {\rm d}u\,\frac{2u-t}{2u}F_{\rm UV}(u,t)\right],
\qquad t=q^2.\label{scalvacpol_eq}
\end{eqnarray}
Using Eq.~(\ref{fuveq}) with $s=\Lambda^2$ provides a simple relation
between the vacuum polarization and the UV-channel function
\begin{eqnarray}
\Pi_{\rm S}(q^2)=
\frac{\Lambda^2}{4\pi^2}
\frac{1}{\lambda}\left[F_{\rm UV}(\Lambda^2,q^2)-1\right],
\label{scalvacpol_eq2}
\end{eqnarray}
which is different from the functional form proposed 
in Ref.~\cite{kotaya93}.

The integrals equations for $F_{\rm IR}$ and $F_{\rm UV}$
are equivalent to two second order differential equations
with four appropriate boundary conditions.
We get for the IR channel:
\begin{eqnarray}
s^2 \frac{{\rm d}^2}{{\rm d}s^2}F_{\rm IR}(s,t)
+2s\frac{{\rm d}}{{\rm d}s}F_{\rm IR}(s,t)+
\frac{\lambda}{2t} s F_{\rm IR}(s,t)=0,
\label{eq:IR}
\end{eqnarray}
and for the UV channel
\begin{eqnarray}
s^2 \frac{{\rm d}^2}{{\rm d}s^2}F_{\rm UV}(s,t)+
2s\frac{{\rm d}}{{\rm d}s}F_{\rm UV}(s,t)+
\lambda\frac{2s-t}{2s} F_{\rm UV}(s,t)=0.
\label{eq:UV}
\end{eqnarray}
The infrared and ultraviolet boundary conditions (IRBC), 
respectively, (UVBC)
are
\begin{eqnarray}
\left[s^2\frac{{\rm d}}{{\rm d}s}F_{\rm IR}(s,t)\right]\Biggr|_{s=0}=0,\qquad
\left[F_{\rm UV}+s\frac{{\rm d}}{{\rm d}s}F_{\rm UV}\right]
\Biggr|_{s=\Lambda^2}=1.
\label{BC}
\end{eqnarray}
Moreover we get a continuity and differentiability equation
at $s=t$
\begin{eqnarray}
F_{\rm IR}(t,t)=F_{\rm UV}(t,t),\qquad 
\frac{{\rm d}}{{\rm d}s}F_{\rm IR}(s,t)\Biggr|_{s=t}=
\frac{{\rm d}}{{\rm d}s}F_{\rm UV}(s,t)\Biggr|_{s=t}.\label{contdiff}
\end{eqnarray}
The differential equations can be solved straightforwardly.
The equation for $F_{\rm IR}$ can be written as a Bessel equation, and
the equation for $F_{\rm UV}$ as a modified Bessel equation.
The general solutions of the differential equations are
\begin{eqnarray}
F_{\rm IR}(s,t)&=&c_3(t/\Lambda^2,\omega) 
\left(\frac{t}{s}\right)^{1/2} J_1\left(\sqrt{2\lambda s/t }\right)+
c_4(t/\Lambda^2,\omega) 
\left(\frac{t}{s}\right)^{1/2}  Y_1\left(\sqrt{2\lambda s/t}\right),
\label{fir}\\
F_{\rm UV}(s,t)&=&c_1(t/\Lambda^2,\omega) \left(\frac{t}{s}\right)^{1/2} 
I_{-\omega}\left(\sqrt{2\lambda t/s}\right)+
c_2(t/\Lambda^2,\omega) \left(\frac{t}{s}\right)^{1/2} 
I_{\omega}\left(\sqrt{2\lambda t/s}\right),\label{fuv}
\end{eqnarray}
with $J_1$ and $Y_1$ the Bessel functions
of first and second kind,  respectively, and where
$I_{\pm\omega}$ are modified Bessel functions and  
$\omega=\sqrt{1-4\lambda}$, see also Eq.~(\ref{omega}). We note that since 
Eqs.~(\ref{eq:IR}), (\ref{eq:UV}) are scale invariant their solutions
are functions of the ratio $s/t$ and the scale invariance is violated by 
the UV boundary condition (\ref{BC}) only.

The IRBC for the IR channel requires $c_4(t/\Lambda^2,\omega)=0$, since the
Bessel function $Y_n(z)$ is irregular at $z=0$,
the other coefficients are fixed by the remaining three boundary
conditions and the solutions are
\begin{eqnarray}
c_1(t/\Lambda^2,\omega)&=&\frac{\pi\gamma(\omega)}{2\sin\omega\pi}
Z^{-1}(t/\Lambda^2,\omega),\\
c_2(t/\Lambda^2,\omega)&=&c_1(t/\Lambda^2,-\omega),\\
c_3(t/\Lambda^2,\omega)&=&Z^{-1}(t/\Lambda^2,\omega),
\end{eqnarray}
where
\begin{eqnarray}
Z(t/\Lambda^2,\omega)\equiv \frac{\pi}{2\sin\omega\pi}
\left[\gamma(\omega)G(t/\Lambda^2,-\omega)
-\gamma(-\omega)G(t/\Lambda^2,\omega) \right],\label{Zdef}
\end{eqnarray}
and
\begin{eqnarray}
\gamma(\omega)&\equiv&\sqrt{2\lambda}\left[
J_1(\sqrt{2\lambda})I_\omega^\prime(\sqrt{2\lambda})
+J_1^\prime(\sqrt{2\lambda})I_\omega(\sqrt{2\lambda})\right],\label{gammaeq}\\
G(t/\Lambda^2,\omega)&\equiv&\frac{1}{2}\sqrt{\frac{t}{\Lambda^2}}
\left[I_\omega\left(\sqrt{2\lambda t/\Lambda^2}\right)-
\sqrt{2\lambda t/\Lambda^2}I_\omega^\prime
\left(\sqrt{2\lambda t/\Lambda^2}\right)\right].
\label{Geq}
\end{eqnarray}
Summarizing, the solution of scalar vertex in terms of channel functions is
\begin{eqnarray}
F_{\rm IR}(p^2,q^2)&=&Z^{-1}\left(\frac{q^2}{\Lambda^2},\omega\right) 
\left(\frac{q^2}{p^2}\right)^{1/2}
J_1\left(\sqrt{\frac{2\lambda p^2}{q^2} }\right),
\label{fir2}\\
F_{\rm UV}(p^2,q^2)&=&
\frac{\pi}{2\sin \omega\pi} Z^{-1}\left(\frac{q^2}{\Lambda^2},\omega\right)
\left(\frac{q^2}{p^2}\right)^{1/2}\nonumber\\
&\times&\left[ 
\gamma(\omega)I_{-\omega}\left(\sqrt{\frac{2\lambda q^2}{p^2}}\right)
-\gamma(-\omega)I_{\omega}\left(\sqrt{\frac{2\lambda q^2}{p^2}}\right)\right].
\label{fuv2}
\end{eqnarray}

It is easy to verify that at zero boson momentum one gets
\begin{equation}
{\Gamma_{\rm S}}(p,p)\equiv F_{\rm UV}(p^2,q^2=0)
=\frac{2}{1+\omega}\left(\frac{p^2}
{\Lambda^2}\right)^{-(1-\omega)/2},\label{GSqzero}
\end{equation}
which coincides with the zero transfer vertex of Refs.~\cite{kotaya93,aptewij91}.
In the limit of pure NJL model, $\alpha_0\to0\quad (\omega\to1)$, the 
vertex is equal, of course, to the bare vertex, 
$\Gamma_S=1=F_{\rm UV}=F_{\rm IR}$. To study the vertex at critical 
gauge coupling $\alpha_0=\alpha_c\quad (\omega=0)$ we expand the Bessel 
functions in small $\omega$ using the following property of the 
modified Bessel functions:
\begin{eqnarray}
I_\omega(x(\omega))\approx I_0(x(0))-\omega K_0(x(0))+{\cal O}(\omega^2),
\end{eqnarray}
where $x(\omega)\propto\sqrt{1-\omega^2}$.
Then the expressions for $F_{\rm IR}$, $F_{\rm UV}$, 
Eqs.~(\ref{fir2}), (\ref{fuv2}), take the following form at 
$\alpha_0\rightarrow\alpha_c$:
\begin{eqnarray}
F_{\rm IR}(p^2,q^2)&=&Z^{-1}\left(\frac{q^2}{\Lambda^2},0\right) 
\left(\frac{q^2}{p^2}\right)^{1/2}J_1\left(\sqrt{\frac{p^2}{2q^2} }\right),\\
\label{vertirwzero}
F_{\rm UV}(p^2,q^2)&=&
Z^{-1}\left(\frac{q^2}{\Lambda^2},0\right)\left(\frac{q^2}{p^2}\right)^{1/2}
\left[\epsilon_1K_0\left(\sqrt{\frac{q^2}{2p^2}}\right)
-\epsilon_2I_0\left(\sqrt{\frac{q^2}{2p^2}}\right)\right],
\label{vertuvwzero}
\end{eqnarray}
where 
\begin{eqnarray}
Z^{-1}\left(\frac{q^2}{\Lambda^2},0\right)&=&
\frac{1}{2}\left(\frac{q^2}{\Lambda^2}\right)^{1/2}\left[
\epsilon_1 K_0(x)
+\epsilon_1 x K_1(x)-\epsilon_2 I_0(x)-\epsilon_2 x I_0^\prime(x)\right],\\
x&=&\sqrt{q^2/2\Lambda^2},\nonumber
\end{eqnarray}
\begin{eqnarray}
\epsilon_1&=&\sqrt{1/2}(
J_1(\sqrt{1/2})I_0^\prime(\sqrt{1/2})+J_1^\prime(\sqrt{1/2})I_0(\sqrt{1/2})),\\
\epsilon_2&=&\sqrt{1/2}(
J_1(\sqrt{1/2})K_0^\prime(\sqrt{1/2})+J_1^\prime(\sqrt{1/2})K_0(\sqrt{1/2})),
\end{eqnarray}
and $K_i$ is the modified  Bessel function of the third kind.
Note that when we expand the UV channel function, ($p^2\gg q^2$),
Eq.~(\ref{vertuvwzero}), we get
\begin{eqnarray}
F_{\rm UV}(p^2,q^2)=2\left(\frac{p^2}{\Lambda^2}\right)^{-1/2}
\left[\frac{\epsilon_3-2+\ln(p^2/q^2)}{\epsilon_3
-\ln(q^2/\Lambda^2)} +{\cal O}\left(q^2/p^2\ln(q^2/p^2)\right)\right],
\label{FUVkondo}
\end{eqnarray}
where
\begin{eqnarray}
\epsilon_3=2(1-\gamma)+3\ln 2-2\frac{\epsilon_2}{\epsilon_1}\approx 3.2
\end{eqnarray}
and $\gamma$ is the Euler gamma.
The expression (\ref{FUVkondo}) is of the same form as 
obtained in Ref.~\cite{kotaya93} (their formula (2.88)).

\section{The Scalar Propagator}
\label{scal_prop}

In the previous section we obtained an analytical expression for the
scalar vertex by assuming that the vertex only weakly depends on the angle
between
boson momentum and fermion momentum.
The expression for the scalar vacuum polarization in such an approximation
takes the form of Eq.~(\ref{scalvacpol_eq}).
The main object of investigation is scalar compositeness near the 
critical line.
In the neighborhood of the critical line,
the tendency of fermion-antifermion pairs to form bound states under
the influence of strong attractive four-fermion forces becomes apparent.

Since we are in the symmetric phase of the GNJL model
(there is no dynamical mass), the only important variable is the scalar boson 
momentum over cutoff $t=q^2/\Lambda^2$.
Equation~(\ref{scalvacpol_eq}) shows that the vacuum polarization depends only 
on the UV channel with fermion momentum at $\Lambda^2$, 
Eq.~(\ref{scalvacpol_eq2}).
Substituting the expressions obtained for the vertex function
Eq.~(\ref{fuv}) in the equation for the vacuum polarization
Eq.~(\ref{scalvacpol_eq2}), we obtain 
\begin{eqnarray}
\Pi_{\rm S}(q^2)=
\frac{\Lambda^2}{4\pi^2}\frac{1}{\lambda}
\left[\frac{\gamma(\omega)H(q^2/\Lambda^2,-\omega)
-\gamma(-\omega)H(q^2/\Lambda^2,\omega)}{\gamma(\omega)
G(q^2/\Lambda^2,-\omega)
-\gamma(-\omega)G(q^2/\Lambda^2,\omega)} \right],
\label{vacas1}
\end{eqnarray}
where $\gamma$ and $G$ are given by Eqs.~(\ref{gammaeq}) and (\ref{Geq}), 
and where
\begin{eqnarray}
H(t/\Lambda^2,\omega)&\equiv&\frac{1}{2}\sqrt{\frac{t}{\Lambda^2}}
\left[I_\omega\left(\sqrt{2\lambda t/\Lambda^2}\right)+
\sqrt{2\lambda t/\Lambda^2}
I_\omega^\prime\left(\sqrt{2\lambda t/\Lambda^2}\right)\right].
\label{Heq}
\end{eqnarray}
This expression is valid along the entire critical curve in the symmetric
phase. 
Note that the vacuum polarization is symmetric in $\omega$ which means
that we can analytically continuate it to values $\alpha_0>\alpha_c$.

Equation~(\ref{vacas1}) is a rather complicated expression, and we 
will first investigate some specific limits:
\begin{enumerate}
\item[(A)]{
The pure NJL limit, where the gauge interaction is turned off,
i.e., the case where $\alpha_0=0$, thus $\omega=1$.}
\item[(B)]{The behavior at the critical gauge coupling
$\alpha_0=\alpha_c$, thus $\omega=0$.}
\item[(C)]{The behavior of $\Pi_{\rm S}$ for $\alpha_0>\alpha_c$, $\omega=i\nu$,
$\nu=\sqrt{\alpha_0/\alpha_c-1}$, i.e., analytic
continuation across the critical curve at $\alpha_0=\alpha_c$.}
\item[(D)]{Asymptotic behavior of $\Pi_{\rm S}$, the infrared behavior 
$q^2/\Lambda^2\ll 1$ in such a manner, so that
$(q^2/\Lambda^2)^\omega \gg q^2/\Lambda^2$.}
\end{enumerate}
\subsection{The pure NJL limit}
The pure Nambu--Jona-Lasinio limit is the case where the gauge interaction
is completely turned off, possible bound states in such a model are 
purely due to the four-fermion interaction.
So $\alpha_0\rightarrow 0$, and therefore $\omega\rightarrow 1$.
The scalar vertex at $\alpha_0=0$ is just equal to the bare vertex, 
${\Gamma_{\rm S}}=1=F_{\rm UV}=F_{\rm IR}$.
The pure NJL limit can be correctly obtained from Eq.~(\ref{vacas1}) by making
expansions of the Bessel functions keeping sufficient number of terms, 
and then performing an expansion in $(1-\omega)$ to zeroth order.
Hence with $\alpha_0=0$, $\omega=1$ in Eq.~(\ref{vacas1}), we get
\begin{eqnarray}
\Pi_{\rm S}(q^2)= \frac{\Lambda^2}{4\pi^2}
\left[1+\frac{q^2}{2\Lambda^2}\ln \left(\frac{q^2}{\Lambda^2}\right)
-\frac{3q^2}{4\Lambda^2}\right].\label{vacas_njl}
\end{eqnarray}
The log term is a consequence of the ``hard'' fermion loop
which also ruins the renormalizability of the pure NJL model.
\subsection{The $\sigma$ boson at critical gauge coupling}
At critical gauge coupling $\alpha_0=\alpha_c$ ($\omega=0$) is the onset 
of the scalar compositeness origining purely from electromagnetic forces. 
The expression for $F_{\rm UV}$, Eq.~(\ref{vertuvwzero}), has the form 
at $\alpha_0=\alpha_c$
\begin{eqnarray}
F_{\rm UV}(p^2,q^2)&=&2\left(\frac{p^2}{\Lambda^2}\right)^{-1/2}
\left[\frac{\epsilon_2 I_0(y)-\epsilon_1  K_0(y)}{
\epsilon_2 I_0(x)-\epsilon_2 x I_1(x)
-\epsilon_1 K_0(x)-\epsilon_1 x K_1(x)}\right], 
\label{fuvwzero}
\end{eqnarray}
where $x=\sqrt{q^2 / 2\Lambda^2}$ and $y=\sqrt{q^2/2 p^2}$.
This gives 
\begin{eqnarray}
\Pi_{\rm S}(q^2)=\frac{\Lambda^2}{4\pi^2}\left[
4+\frac{8\left[\epsilon_2 x I_1(x)+\epsilon_1 x K_1(x)\right]}{
\epsilon_2 I_0(x)-\epsilon_2 x I_1(x)-\epsilon_1 K_0(x)-\epsilon_1 x K_1(x)}
\right]. 
\label{alpcrit}
\end{eqnarray}
\subsection{Analytic continuation across the critical curve}
Since the expression for the scalar vacuum polarization is symmetric in 
$\omega$, Eq.~(\ref{vacas1}), it can be analytically continuated to the values
 $\alpha_0> \alpha_c$.
This holds in replacing $\omega$ by $i\nu$ in Eq.~(\ref{vacas1}), where
\begin{eqnarray}
\nu=\sqrt{\alpha_0/\alpha_c-1}.
\end{eqnarray}
\subsection{Asymptotic behavior of the $\sigma$ boson vacuum polarization}
The asymptotic behavior $0<\omega< 1$ so that
$(q^2/\Lambda^2)^\omega \gg q^2/\Lambda^2$ can be obtained 
by considering first the $q^2\ll \Lambda^2$ limit of $Z$, Eq.~(\ref{Zdef}): 
\begin{eqnarray}
Z\approx \frac{\pi}{2\sin \omega \pi}
\sqrt{h(\omega)h(-\omega)(1-\omega^2)} 
\left(\frac{q^2}{\Lambda^2}\right)^{1/2}
\sinh \left[\frac{\omega}{2}\ln\left(\frac{\Lambda^2}{ q^2}\right)
+\delta(\omega)\right],
\end{eqnarray}
where
\begin{eqnarray}
\delta(\omega)\equiv
\frac{1}{2}\ln\frac{ h(\omega)(1+\omega)}{h(-\omega)(1-\omega)}
-\omega\ln \sqrt{2\lambda},\qquad
h(\omega)\equiv \frac{\gamma(\omega) 2^\omega}{\Gamma(1-\omega)}.
\end{eqnarray}
Then the UV-channel function with fermion momentum $p^2=\Lambda^2$
can be expressed, in this limit, as
\begin{eqnarray}
F_{\rm UV}(\Lambda^2,q^2)\approx\frac{2}{1+\omega}
+\frac{2\omega}{1-\omega^2}(1-\coth y),
\qquad y={\omega\over2}\ln\left(\frac{\Lambda^2}{q^2}\right)+\delta(\omega).
\end{eqnarray}
Hence
\begin{eqnarray}
\Pi_{\rm S}(q^2)&\approx&
\frac{\Lambda^2}{4\pi^2}
\left[\frac{1}{g_c(\omega)}+\frac{8\omega}{(1-\omega^2)^2}(1-\coth y)
\right],\qquad q^2\ll \Lambda^2.
\label{vacasym}
\end{eqnarray}
However, the above expression does not reproduce the correct leading term of 
the NJL limit ($\omega\rightarrow 1$), for that we should use the expression
(\ref{vacas1}). In order to obtain such an expression which contains properly
the pure NJL limit, we have to keep more terms in the expansion of $Z$.
We then get for the scalar vacuum polarization
\begin{eqnarray}
\Pi_{\rm S}(q^2)&\approx&
\frac{\Lambda^2}{4\pi^2}\left[\frac{1}{g_c(\omega)}
-B(\omega)\left(\frac{q^2}{\Lambda^2}\right)^\omega\hspace{-2mm}
+A(\omega) \frac{q^2}{\Lambda^2}
+{\cal O}\left( (q^2/\Lambda^2)^{2\omega}\right)
+{\cal O}\left( (q^2/\Lambda^2)^{1+\omega}\right)
\right],
\label{vacasym2}
\end{eqnarray}
where
\begin{eqnarray}
A(\omega)\equiv \frac{1}{2 g_c(\omega)(1-\omega)},
\qquad B(\omega)=\frac{16 \omega}{(1-\omega^2)^2}
\frac{\gamma(-\omega)}{\gamma(\omega)} 
\frac{\Gamma(2-\omega)}{\Gamma(2+\omega)}
\left(\frac{\lambda}{2}\right)^\omega.
\label{Bw}
\end{eqnarray}
For Eq.~(\ref{vacasym2}) to be valid, it is assumed that $\omega> 1/2$.
It is straightforward to check that this expression satisfies
the NJL limit ($\omega\rightarrow 1$).

This is a suitable point to refer to Appendix~\ref{Chebyshev}
for an analysis of the reliability of Eq.~(\ref{vacasym2}), 
where it is shown that the leading and next-to-leading terms 
in $q^2/\Lambda^2$ of $\Pi_{\rm S}$ (i.e., the first two terms on 
the right-hand side of 
Eq.~(\ref{vacasym2})) are indeed correctly obtained by our approximation, 
see Eq.~(\ref{error_expr}).

The four specific limits of the scalar vacuum polarization described above
are very useful for studying the resonance structure of the
bound states which will be done in Sec.~\ref{res_near_crit}, and
for the study of the CPT which will be done in Sec.~\ref{sec_cpt}.
But first we shall compare our result for $\Delta_{\rm S}$ with that 
one obtained by other authors in the next section.

\section{Comparison with earlier works}
\label{sec_review}

In this section we discuss the earlier work of Appelquist,
Terning, and Wijewardhana \cite{aptewij91} and related work based on that
by Kondo, Tanabashi, and Yamawaki \cite{kotaya93} on the
scalar composites in the GNJL model.
The method used by these authors to solve the coupled set of
scalar vertex and scalar vacuum polarization is the following.
They consider a Taylor-series expansion about $q=0$ of the
scalar vacuum polarization $\Pi_{\rm S}(q^2)$.
In the ladder approximation, such a series has the property that
the $n$th derivative can be written as
\begin{eqnarray}
\left(\frac{{\rm \partial}}{\partial q}\right)^n\Pi_{\rm S}(q^2)
\sim \int_\Lambda \frac{{\rm d}^4k}{(2\pi)^4}\,
\sum_{m=0}^{n-1} C_m
{\rm Tr}\left[\left(\frac{\partial^m}{\partial q^m}{\Gamma_{\rm S}}(k+q,k)\right)
\frac{1}{\hat k}{\Gamma_{\rm S}}(k,k+q)
\frac{\partial^{n-m}}{\partial q^{n-m}}\frac{1}{\hat k-\hat q}\right].
\end{eqnarray}
Their basic assumption is then
that derivatives of the scalar vertex ${\Gamma_{\rm S}}$
can be neglected with respect to ${\Gamma_{\rm S}}$ for small $\alpha_0$,
\begin{eqnarray}
\frac{\partial^n{\Gamma_{\rm S}}(k+q,k)}{\partial q^n}
\Biggr|_{q=0}\sim\frac{1}{k^n}
\frac{\alpha_0}{4\alpha_c}{\Gamma_{\rm S}}(k,k).\label{dervscalvert}
\end{eqnarray}
Subsequently the Taylor series is resummed using the assumption stated above
to obtain
\begin{eqnarray}
\Pi_{\rm S}(q^2)\sim
\int_\Lambda \frac{{\rm d}^4k}{(2\pi)^4}\,
{\rm Tr}\left[{\Gamma_{\rm S}}(k,k) \frac{1}{\hat k}{\Gamma_{\rm S}}(k,k)
\frac{1}{\hat k-\hat q}\right],
\end{eqnarray}
which yields in terms of Euclidean momentum,
\begin{eqnarray}
\Pi_{\rm S}(q^2)=
\frac{\Lambda^2}{4\pi^2}
\left[\frac{1}{g_c(\omega)}-b(\omega)\left(\frac{q^2}{\Lambda^2}\right)^\omega
+a(\omega)\frac{q^2}{\Lambda^2}\right],
\label{appelres}
\end{eqnarray}
where
\begin{eqnarray}
a(\omega)=\frac{1}{2g_c(\omega)(1-\omega)},
\qquad b(\omega)=\frac{1}{g_c(\omega)\omega(1-\omega^2)},\qquad
g_c=\frac{(1+\omega)^2}{4}.\label{aandb}
\end{eqnarray}
How does their result compare to ours?
From our expression for the asymptotic behavior of $\Pi_{\rm S}$,
Eq.~(\ref{vacasym2}), and the result obtained in
Refs.~\cite{aptewij91} and
\cite{kotaya93}, Eq.~(\ref{appelres}),
the leading power of momentum is the same, namely, $(q^2/\Lambda^2)^\omega$.
However, the $\omega$-dependent factors in front of the leading 
and next-to-leading powers are
different. At the same time the pure NJL limit is obtained correctly 
in Refs.~\cite{aptewij91} and \cite{kotaya93}.
The differences are rather small for values of $\omega$ close to 1, i.e.
the coefficients are
\begin{eqnarray}
A(\omega)=a(\omega),\qquad 
B(\omega)=b(\omega)\left[1+{\cal O}((1-\omega)^2)\right].
\end{eqnarray}
Hence the approach to the NJL point $\omega=1$ of both approximations 
is equal. However, for smaller values of $\omega$ the coefficient $B$ 
obtained in the present paper starts to deviate from the one obtained 
in Ref.~\cite{aptewij91}. Then for such values of $\omega$, 
$b(\omega)>B(\omega)$.

What is the origin of this difference?
The first point is that the expression derived in Refs.~\cite{aptewij91} and
\cite{kotaya93} is valid for $\alpha_0$ not too large.
Secondly, from their answer Eq.~(\ref{appelres}) it is clear that a Taylor
series of the scalar propagator about $q=0$ is not well defined due to the
noninteger power behavior for values $0<\omega<1$.
This is reflected also in their assumption regarding the derivatives
of the scalar vertex at $q=0$, Eq.~(\ref{dervscalvert}).
The expression for the scalar vertex obtained in
Sec.~\ref{sec_scal_vertex} shows that in general, for $0<\omega<1$,
the assumption (\ref{dervscalvert}) is not true.
Such derivatives of ${\Gamma_{\rm S}}$ are singular at $q=0$ due to the fact
that they depend on noninteger powers of $q$, which can be seen
from Eq.~(\ref{fuv}).

The scalar vertex for small $q\ll p$ is of the form
\begin{eqnarray}
{\Gamma_{\rm S}}(p+q,p)\sim F_{\rm UV}(p^2,q^2)
&\approx&
\left(\frac{p^2}{\Lambda^2}\right)^{-1/2+\omega/2}\Biggr\{\frac{2}{(1+\omega)}
+\frac{q^2}{p^2}\left[\frac{1}{4}+\frac{(1-\omega)}{4(1+\omega)}
\frac{p^2}{\Lambda^2}\right]
\nonumber\\
&-&2\left(\frac{q^2}{p^2}\right)^\omega 
\frac{\gamma(-\omega)}{\gamma(\omega)}
\frac{\Gamma(1-\omega)}{\Gamma(2+\omega)}\left[1
-\frac{(1-\omega)}{(1+\omega)}
\left(\frac{p^2}{\Lambda^2}\right)^\omega\right]
\left(\frac{\lambda}{2}\right)^\omega\nonumber\\
&+&{\cal O}\left((q^2/p^2)^\omega (q^2/\Lambda^2)^\omega\right)
+{\cal O}\left((q^2/p^2)^{1+\omega}\right)+\cdots\Biggr\}\label{fuv_asym},
\end{eqnarray}
which is consistent with
Eq.~(\ref{vacasym2}) for $p^2=\Lambda^2$, because of 
Eq.~(\ref{scalvacpol_eq2}).
Of course at $q=0$ our result coincides with that of the other authors.
But for nonzero $q$ this expression clearly shows that for $0<\omega<1$ the 
vertex contains noninteger powers of $q$. Hence the assumption made in 
Refs.~\cite{kotaya93,aptewij91} is not true in general, since higher
derivatives of the scalar vertex with respect to $q$ are singular at $q=0$.

In what follows, we discuss the results from the viewpoint of 
the renormalizability of the GNJL model.
The renormalization is performed by a suitable redefinition of the 
composite or auxilary fields $\sigma$ and $\pi$:
\begin{eqnarray}
\sigma_R\equiv \left[Z_\sigma^{(\mu)}\right]^{-1}\sigma,\qquad
\pi_R\equiv\left[Z_\pi^{(\mu)}\right]^{-1}\pi.\label{ren_aux}
\end{eqnarray}
The renormalization factors of the scalar and pseudoscalar fields
$Z_\sigma^{(\mu)}$, respectively, $Z_\pi^{(\mu)}$ can be chosen to coincide
both in the symmetric and broken phase,
and the renormalized auxilary fields $\sigma_R$ and $\pi_R $ define
the renormalized scalar propagator
\begin{eqnarray}
\Delta_{\rm S}^{R}(q)=\left[Z_\sigma^{(\mu)}\right]^{-2} \Delta_{\rm S}(q)
\end{eqnarray}
and the renormalized scalar vertex
\begin{eqnarray}
\Gamma_{\rm S}^R(p+q,p)= Z_\sigma^{(\mu)} {\Gamma_{\rm S}}(p+q,p).
\end{eqnarray}
In order to renormalize the scalar propagator and Yukawa vertex 
simultaneously (see also Eqs.~(\ref{vacasym2}) and (\ref{fuv_asym})), 
the wave function renormalization factor at some arbitrary renormalization 
scale $\mu$ should be of the form
\begin{eqnarray}
Z_\sigma^{(\mu)}\propto 
\left(\frac{\mu^2}{\Lambda^2}\right)^{(1-\omega)/2}.
\label{Zscal}
\end{eqnarray}
Freedom in the choice of renormalization scheme allows us to
take the factor $Z$ defined in Eq.~(\ref{Zdef}) as the wave function 
renormalization factor, since
\begin{eqnarray} 
Z_\sigma^{(\mu)}=Z(\mu^2/\Lambda^2,\omega)\approx 
\frac{\pi}{2\sin \omega\pi}\frac{\gamma(\omega)}{2}
\frac{(1+\omega)}{\Gamma(1-\omega)}
\left(\frac{\lambda}{2}\right)^{-\omega/2}
\left(\frac{\mu^2}{\Lambda^2}\right)^{(1-\omega)/2}.
\end{eqnarray}
Hence it follows that four-fermion scattering amplitudes, for
instance one-scalar exchange amplitudes are
renormalization group (RG) invariant, i.e.
\begin{eqnarray}
\Gamma_{\rm S}^R(p_1+q,p_1) \Delta_{\rm S}^{R}(q) 
\Gamma_{\rm S}^R(p_2,p_2+q)\sim
{\Gamma_{\rm S}}(p_1+q,p_1)\Delta_{\rm S}(q){\Gamma_{\rm S}}(p_2,p_2+q).
\label{RGscat}
\end{eqnarray}
Consider the case where $p_1^2,\,p_2^2 \gg q^2$,
so that the scalar vertices are described by the UV channels,
and suppose that we are sufficiently close to the critical line
\begin{eqnarray}
\Delta g_0\equiv g_c-g_0 \ll \left(\frac{q^2}{\Lambda^2}\right)^\omega
\Longrightarrow \Delta g_0\sim \left(\frac{\mu^2}{\Lambda^2}\right)^\omega,
\qquad \mu^2\ll q^2,
\end{eqnarray}
then from Eqs.~(\ref{delsinv}) and (\ref{vacasym2}) the scalar propagator has the 
asymptotic behavior
\begin{eqnarray}
\Delta_{\rm S}(q)\approx -\frac{4\pi^2}{\Lambda^2}\frac{1}{B(\omega)}
\left(\frac{\Lambda^2}{q^2}\right)^\omega.
\end{eqnarray}
Such a specific power-law behavior for the scalar propagator 
is essential for the renormalizability of the GNJL model as is shown
in Refs.~\cite{gusmir,miransky,kotaya93}.

Thus, from Eq.~(\ref{Zscal}) and Eq.~(\ref{fuv_asym}), we get
\begin{eqnarray}
\Delta_{\rm S}^{R}(q)&\propto&
-\frac{1}{\mu^2}\left(\frac{\mu^2}{q^2}\right)^\omega,\\
F_{\rm UV}^R\left(p^2,q^2\right)&\propto&
\frac{2}{(1+\omega)}\left(\frac{p^2}{\mu^2}\right)^{-(1-\omega)/2 }
\left[1-2\left(\frac{q^2}{p^2}\right)^\omega
\frac{\gamma(-\omega)}{\gamma(\omega)}
\frac{\Gamma(1-\omega)}{\Gamma(2+\omega)}
\left(\frac{\lambda}{2}\right)^\omega \right].
\end{eqnarray}
With these expressions, it is straightforward to check that
Eq.~(\ref{RGscat}) is indeed independent of $\Lambda$ and $\mu$.
Hence the renormalization of the auxilary fields 
$\sigma$ and $\pi$, Eq.~(\ref{ren_aux}), simultaneously renormalizes 
the Yukawa vertex and the scalar propagator.

\section{Light Scalar Resonances near criticality}\label{res_near_crit}

In this section we discuss the behavior of the scalar propagator near the
critical line in the symmetric phase $g_0\leq g_c$.
In the symmetric phase the scalar and pseudoscalar composites, the $\sigma$
and $\pi$ bosons are degenerate. Near the critical curve, a combination of 
strong four-fermion coupling and gauge coupling will tend to bind fermions 
and antifermions into these scalar composites.
Since the chiral symmetry is unbroken the $\sigma$ and $\pi$ bosons decay
to massless fermions and antifermions.
Hence the scalar composites are resonances which are described by a complex
pole in their propagators. The complex pole determines the mass and the
width of the resonances. In what follows, we redo the computation of the 
complex poles of the $\sigma$ boson which was performed by Appelquist et al. 
in Ref.~\cite{aptewij91} using the expression for 
$\Delta_{\rm S}$, Eq.~(\ref{vacas1}),
obtained with the two-channel approximation of the Yukawa vertex.
The expressions obtained in Sec.~\ref{scal_prop} for $\Pi_{\rm S}(p^2)$
in various regimes are rotated back to Minkowski momentum
$p^2\rightarrow p^2_M \exp(-i\pi)$.
Then the complex poles are given by
\begin{eqnarray}
p_M^2=p_0^2\exp{(-i\theta)},\qquad
\Delta_{\rm S}^{-1}(p_M)=
-\frac{\Lambda^2}{4\pi^2g_0}+\Pi_{\rm S}(p_0^2\exp{(-i\theta)})=0.
\end{eqnarray}
We can also parametrize the location of a pole by a mass and a width, i.e.,
$p_0^2\exp{(-i\theta)}=\left[M_\sigma-(i/2)\Gamma_\sigma\right]^2$,
which yields
\begin{eqnarray}
M_\sigma=p_0\left[\frac{1+\cos\theta}{2}\right]^{1/2},\qquad
\frac{\Gamma_\sigma}{M_\sigma}=\frac{2\sin\theta}{1+\cos \theta}.
\end{eqnarray}
If $\theta$ is small, then $\Gamma_\sigma/M_\sigma\approx \theta$.

Near the Nambu--Jona-Lasinio point ($\alpha_0=0$) our expression for 
the vacuum polarization coincides with that obtained by Appelquist et al., 
Eq.~(\ref{vacas_njl}),
and we get the following equations for the resonances:
\begin{eqnarray}
\frac{c\Lambda^2}{p_0^2}=
\frac{\ln(\Lambda^2/p_0^2)+3/2}{\cos\theta},
\qquad c=\frac{2(1-g_0)}{g_0},
\end{eqnarray}
and
\begin{eqnarray}
\pi+\theta =\left[\ln(\Lambda^2/p_0^2)+\frac{3}{2}\right]\tan\theta.
\end{eqnarray}
If now $g_0$ is tuned close enough to the critical value $g_c=1$, 
so that $\ln 1/c \gg 1$, the solution is approximately
\begin{eqnarray}
p_0^2\approx\frac{2(1-g_0)}{g_0 \ln\left[g_0/2(1-g_0)\right]}\Lambda^2
\end{eqnarray}
and we find a narrow width
\begin{eqnarray}
\theta \approx \frac{\pi}{\ln\left[g_0/2(1-g_0)\right]}.
\end{eqnarray}
These results are nothing else than the familiar NJL results.
For intermediate values of the gauge coupling,
$0<\alpha_0<\alpha_c$ ($0<\omega<1$),
we assume the poles of $\Delta_{\rm S}$ are small, $p_0/\Lambda\ll 1$, so that
\begin{eqnarray}
\left(p_0/\Lambda\right)^\omega \ll 1. \label{restriction}
\end{eqnarray}
Then, from Eq.~(\ref{vacasym2}) we get the following equation for 
the real part of the pole:
\begin{eqnarray}
0\approx-\frac{1}{g_0}+\frac{1}{g_c}-B(\omega)
\left(\frac{p_0^2}{\Lambda^2}\right)^\omega
\cos\omega(\theta+\pi).
\end{eqnarray}
The equation for the imaginary part reads
\begin{eqnarray}
0\approx \sin\omega(\theta+\pi),
\end{eqnarray}
where $B(\omega)$ is given by Eq.~(\ref{Bw}).

\begin{figure}[h]
\epsfxsize=10cm
\epsffile[120 320 370 600]{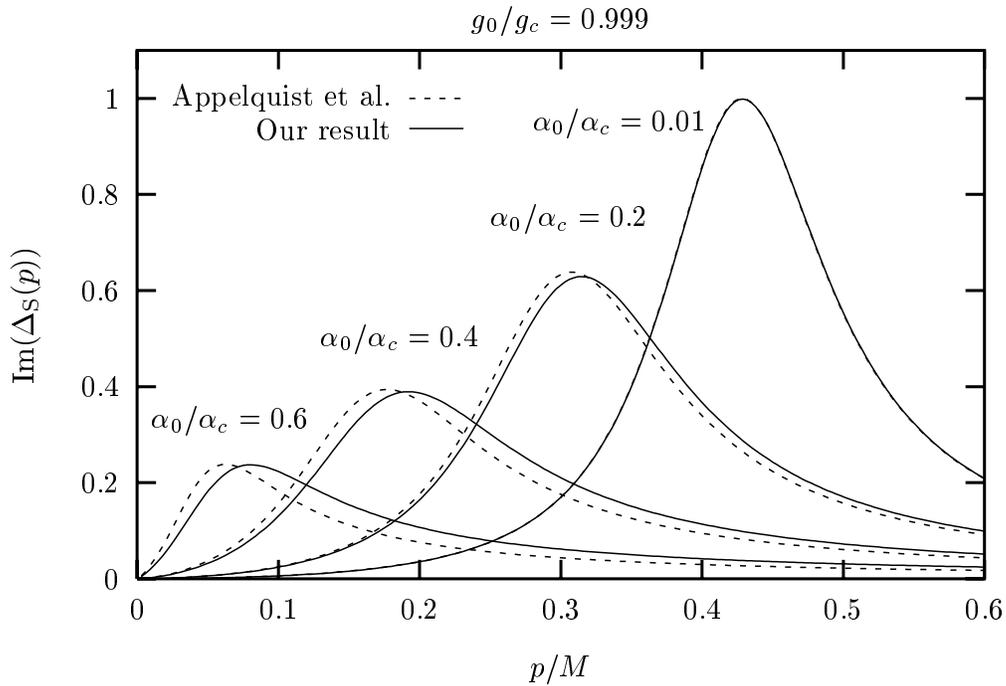}
\caption{A comparison of our result with Appelquist et al.
Graphs of ${\rm Im}(\Delta_{\rm S}(p))$ for different values of
$\alpha_0/\alpha_c$.
The curves are normalized so that the peak of the
$\alpha_0/\alpha_c=0.01$ curve equals 1.}
\label{appelfig}
\end{figure}

The solution is
\begin{eqnarray}
\theta=\frac{\pi(n-\omega)}{\omega},
\end{eqnarray}
and $n$ is odd integer, so that $\cos \omega(\theta+\pi)=-1$,
thus
\begin{eqnarray}
p_0\approx 
\Lambda\left[\frac{(1-g_0/g_c)}{g_0 B(\omega)}\right]^{1/2\omega}.
\label{massinter}
\end{eqnarray}
Hence $\theta$ is only small if $\omega\sim 1$ for $n=1$.
The result obtained in Ref.~\cite{aptewij91}, 
see Eq.~(\ref{aandb}), gives
a mass
\begin{eqnarray}
p_0\approx 
\Lambda\left[\frac{(1-g_0/g_c)}{g_0 b(\omega)}\right]^{1/2\omega},
\qquad b(\omega)=\frac{1}{g_c\omega(1-\omega^2)}.
\end{eqnarray}
The pole obtained in Ref.~\cite{aptewij91} is of the same order as
Eq.~(\ref{massinter}) for values of $\omega$ close to 1
(see also discussion in the previous section).
For more intermediate values of $\omega$ the poles obtained 
in our approximation are somewhat bigger, since 
$b(\omega) > B(\omega)$ for $0<\omega<1$.
The quantitative difference between the
result of Ref.~\cite{aptewij91} and that
obtained in this paper for resonance structures
are visualized in Fig.~\ref{appelfig}.
In Fig.~\ref{appelfig} the imaginary part of $\Delta_{\rm S}$ given by 
Eq.~(\ref{vacas1}) and Eq.~(\ref{delsinv}) 
is plotted versus $p/M$ where the tuning of the four-fermion to the critical
line is $g_0/g_c=0.999$, and $M/\Lambda=(1-g_0/g_c)^{1/2}$.
From Fig.~\ref{appelfig} it is clear that
the position of the peak of the resonant curve is slightly shifted 
to the right in our case at a fixed ratio 
$\alpha_0/\alpha_c \sim {\cal O}(1)$ (intermediate or small values 
of $\omega$), while the width over mass ratio remains comparable.
Near the pure NJL point $\omega\sim 1$ both results coincide.

As was pointed out in the previous section, and following from the restriction
Eq.~(\ref{restriction}), these results are only valid
for $\alpha_0/\alpha_c$ small. For larger values of the gauge coupling, 
$\omega\rightarrow +0$, the widths become larger, and Eq.~(\ref{restriction}) 
is no longer satisfied. This can also be seen in Fig.~\ref{appelfig}.
As the ratio $\alpha_0/\alpha_c$ increases the width increases too,
and the position of the peak becomes more difficult to define.

\section{The Conformal Phase Transition (CPT)}\label{sec_cpt}

In this section we analyze the scalar composites near the critical
gauge coupling $\alpha_0=\alpha_c$, with the purpose of investigating the
conformal phase transition.
The conception of the CPT was introduced and elaborated recently in 
Ref.~\cite{miya97}. It embodies the classification of specific types of 
phase transitions.
The main feature of the CPT is an abrupt change of the spectrum of light 
excitations (composites) as the critical point is crossed, though the
phase transition itself is continuous.
This is connected with the nonperturbative breakdown of the conformal
symmetry by marginal operators ($(\bar\psi\psi)^2+(\bar\psi i\gamma_5\psi)^2$
in the model under consideration), which was illustrated in Ref.~\cite{miya97} 
by a study of the effective potentials in Gross-Neveu and GNJL models 
and quenched QED4.

In the previous section we encountered a no-CPT, $\sigma$-model-like
phase transition for values of $\alpha_0<\alpha_c$ \cite{miya97}.
The masses of light excitations\footnote{The scalar composites are
light resonances.} are continuous functions across the critical
curve; there is no abrupt change in the spectrum of light excitations.
In the broken phase the $\pi$ boson becomes a massless
Nambu-Goldstone boson, while the fermion and $\sigma$ boson acquire
a dynamical mass which is small with respect to the cutoff $\Lambda$ near 
criticality.

The pole at the critical gauge coupling $\alpha_0=\alpha_c$ is determined by 
Eq.~(\ref{alpcrit}), which is rather complicated but 
we only need to study the IR limit, so we assume that the pole is small 
$p_0^2\ll \Lambda^2$.
The infrared limit obtained from Eq.~(\ref{alpcrit}) is
\begin{eqnarray}
\Pi_{\rm S}(q^2)\approx \frac{\Lambda^2}{4\pi^2}\left[
4+\frac{16}{\ln (q^2/\Lambda^2)-\epsilon_3}+{\cal O}\left(
q^2/\Lambda^2\ln (q^2/\Lambda^2)\right)\right].
\end{eqnarray}
\noindent
We then find zeros of $\Delta_S^{-1}(p)$ at
\begin{eqnarray}
0\approx\left(\frac{1}{g_0}-4\right)
+\frac{16(\ln (\Lambda^2/p_0^2)+\epsilon_3)}{
\left(\ln (\Lambda^2/p_0^2)+\epsilon_3\right)^2+(\theta+\pi)^2},
\label{real_ac}
\end{eqnarray}
\begin{eqnarray}
0\approx \theta+\pi.\label{imag_ac}
\end{eqnarray}
Thus $\theta\approx -\pi$, and from Eq.~(\ref{real_ac}) it is clear that 
if $g_0\leq g_c=1/4$ both terms on the
right-hand side are positive, and there is no solution for the pole with
$p_0/\Lambda \ll 1$. 
Hence if there is a pole it will be heavy, i.e., $p_0/\Lambda={\cal O}(1)$.
Therefore at $\alpha_0=\alpha_c$ no light resonances are present in 
the spectrum for $g_0\leq g_c=1/4$.
The imaginary phase $\theta$ approaches $-\pi$ which means the heavy
pole occurs at ``Euclidean'' momentum, a sign of tachyonic states.

The statement above can be made more explicit.
If we analytically continue the scalar propagator to the values
of $\alpha_0>\alpha_c$, then we end up in the ``wrong vacuum''
and we should get tachynonic states. In the broken phase
($\alpha_0>\alpha_c$), a chiral symmetric solution still exists, but
it is unstable. The $\pi$ and $\sigma$ bosons are tachyons for 
such a solution.
The unstable symmetric solution is obtained by analytic continuation
of the solution in the symmetric phase across the critical curve
(at $\alpha_c$).
The scaling law is determined by the UV properties of the theory and therefore
the scaling law of the tachyonic masses is the same as that
of the fermion and $\sigma$-boson mass in the broken phase.

Tachyons are described by imaginary mass $m^2<0$.
This means the scalar propagator must have a real pole for Euclidean
momentum. If the pole $p_0$ is small, $p_0\ll \Lambda$,
we analytically continue Eq.~(\ref{vacasym}) to $\alpha_0>\alpha_c$
by replacing $\omega$ by $i\nu$, 
\begin{eqnarray}
\omega\rightarrow i\nu\equiv i\sqrt{4\lambda-1}.
\end{eqnarray}
We then obtain
\begin{eqnarray}
\Pi_{\rm S}(q^2)=
\frac{\Lambda^2}{\pi^2}\frac{1-\nu^2-2\nu\cot y}{(1+\nu^2)^2},
\qquad y=\frac{\nu}{2}\ln\left(\frac{\Lambda^2}{q^2}\right)+\nu\phi(\nu^2),
\end{eqnarray}
where
\begin{eqnarray}
\phi(\nu^2)&\equiv&\frac{1}{2i\nu}\ln\frac{h(i\nu)(1+i\nu)}{h(-i\nu)(1-i\nu)}
-\ln \sqrt{2\lambda}.
\end{eqnarray}
The tachyonic pole is then given by the zero of the equation
\begin{eqnarray}
-\frac{\Lambda^2}{4\pi^2g_0}+\Pi_{\rm S}(p_0^2)=0,
\end{eqnarray}
which gives
\begin{eqnarray}
\frac{p_0}{\Lambda}=
\exp\left(-\frac{n\pi}{\nu}-\frac{\beta}{\nu}+\phi(\nu^2)\right),
\end{eqnarray}
where $n$ is a positive integer and
\begin{eqnarray}
 \beta=\tan^{-1}\frac{\nu g_0}{g_0-2\lambda(g_0+\lambda)}.
\end{eqnarray}
The tachyon with largest $p_0$ in the physical region 
$p_0< \Lambda$ corresponds to $n=1$.
If we now consider the limit  $\nu\to0\quad (\lambda\to 1/4)$, we get
\begin{eqnarray}
\beta &\approx&
\frac{2\nu g_0}{g_0- 1/4},\quad
\phi(\nu^2)\approx 1-\gamma-\frac{1}{2}\ln 2-\frac{\epsilon_2}{\epsilon_1}
+{\cal O}(\nu^2).
\end{eqnarray}
In this case,
\begin{eqnarray}
\frac{p_0}{\Lambda}\approx 
\exp\left(\frac{2g_0}{1/4-g_0}+\phi(0)\right)
\exp\left(-\frac{\pi}{\sqrt{4\lambda-1}}\right),
\label{tachpole}
\end{eqnarray}
which is proportional to the well-know scaling law of quenched QED.
Thus the scalar propagator giving the
tachyon pole equation, Eq.~(\ref{tachpole}), reproduces the scaling law
with essential singularity, which is another confirmation of the CPT.

\section{Conclusion}\label{conclusion}

In this paper we studied the scalar composites near criticality
in the GNJL model.
We obtained an analytic expression for the scalar propagator
describing the composite states which is valid along the entire
critical curve of the GNJL model.
We presented a method for solving the Yukawa vertex in the GNJL in the
quenched-ladder approximation. The crucial assumption was that such a vertex
depends only weakly on the angle between $\sigma$-boson momentum and
fermion momentum.
The method presented here incorporated the infrared boundary condition
in a more natural way than previous attempts in this direction.
Also the observation that derivatives of the Yukawa vertex
are singular at zero $\sigma$-boson momentum transfer is a warning
that derivative expansions and Taylor series could fail.
Moreover this is reflected by the property of the scalar composites
having noninteger power-law behavior, which means that, although these states 
are tightly bound, they are not pointlike.

The conclusion of the comparison of the method presented here
and work done previously on
the $\sigma$-boson propagator is the following.
Qualitatively the results obtained by Appelquist et al., Kondo et al.
and by our approximation are in agreement. Both methods yield
a renormalizable $\sigma$-boson propagator and Yukawa vertex
near criticality and find light resonances for gauge coupling
$0<\alpha_0<\alpha_c$.
Quantitatively, at a fixed value of the gauge coupling,
the scalar composites computed in the our case
are slightly heavier with comparable width, as is illustrated in 
Fig.~\ref{appelfig}.

In addition, the scalar composites propagator were examined
for values of the coupling near the critical gauge coupling $\alpha_c$.
Near the critical line $\alpha_0=\alpha_c,g_0<{1\over 4}$ the conformal
phase transition is encountered and the spectrum of light excitations 
(resonances) in the symmetric phase disappears.
Moreover the well-known scaling law with essential singularity,
which is characteristic for the CPT, was recovered by analytic continuation
of the $\sigma$-boson propagator across the critical curve at 
$\alpha_0=\alpha_c$.

\acknowledgments{V.P.G. is grateful to the members of the Research Centre for 
Subatomic Structure of Matter, University of Adelaide, Australia, for 
hospitality during his stay there. He acknowledges fruitful discussions 
with Professor A.~Thomas and Dr.~A.~Williams. The work of V.P.G. is supported 
by Swiss National Science Foundation grant No. CEEC/NIS/96--98/7 IP 051219 
and Foundation of Fundamental Researches of Ministry of Sciences of 
Ukraine under grant No. 2.5.1/003.
M.R. thanks the members of the Bogolyubov Institute for Theoretical 
Physics for their hospitality and support during his visit there last 
summer, where the present work was initiated.
We would like to thank Professor V.A.~Miransky for useful and stimulating 
discussions and we acknowledge Professor M.~Winnink, 
Dr.~A.~Williams, and A.H.~Hams for helpful remarks and carefully 
reading the manuscript.} 

\appendix
\section{Analysis of the Chebyshev Expansion}
\label{Chebyshev}

In this appendix we discuss the validity of the zeroth-order 
Chebyshev expansion for the Yukawa vertex function $F_1$ introduced in 
Sec.~\ref{sec_scal_vertex}.
The problem of angular dependence in the SDEs for the Yukawa vertex functions
$F_1$ and $F_2$ is replaced by an infinite set of Chebyshev harmonics.
Subsequently this set is truncated to the lowest order harmonic of $F_1$, 
which is the only harmonic having nonhomogeneous 
ultraviolet boundary conditions because of the presence of the 
(angular independent) inhomogeneous term $1$.

\begin{figure}[h]
\epsfxsize=10cm
\epsffile[-120 100 380 450]{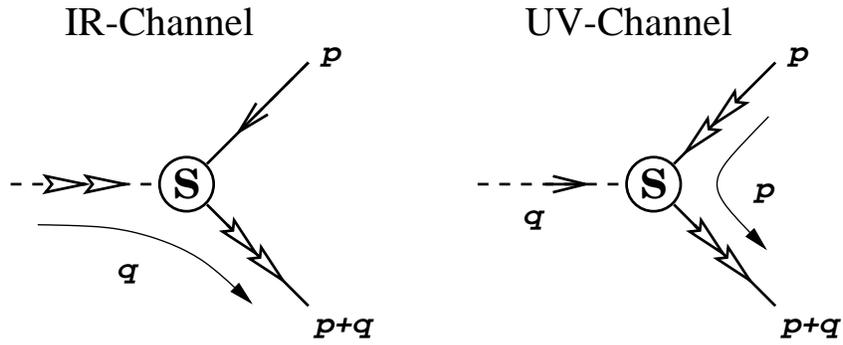}
\caption{The two-channel approximation.}
\label{two_channel_fig}
\end{figure}

As mentioned previously, the method of using expansions 
in terms of Chebyshev polynomials $U_n(x)$ (of the second kind) 
was used before \cite{FGMS,mawa96}. 
These polynomials are orthogonal with respect to the angular integration 
$\int {\rm d} \Omega$. 
In the analysis of BSEs in Ref.~\cite{FGMS} a {\it CP} 
invariant Chebyshev expansion
was used, which has the nice property of keeping only even terms in the 
expansion. 
However, we use a slightly different expansion (not explicitly {\it CP} invariant)
which has the disadvantage of also including odd terms in the Chebyshev 
expansion, but the advantages that the integral equation for the zeroth 
order harmonic is more friendly and the zeroth order harmonic 
coincides with both the large fermion momentum limit ($p^2\gg q^2$) as well 
the large boson-momentum limit ($q^2\gg p^2$) of the Yukawa vertex, see 
Fig.~\ref{two_channel_fig}.

Thus the vertex functions satisfying the SDEs (\ref{sde_vertfies})
are expanded in the angle between fermion momentum $p$ 
and scalar boson $q$, i.e., $p\cdot q$, in the following way:
\begin{eqnarray}
F_1(p+q,p)&=&\sum_{n=0}^\infty f_n(p^2,q^2) U_n(\cos\alpha),\qquad
F_2(p+q,p)=\sum_{n=0}^\infty g_n(p^2,q^2) U_n(\cos\alpha),\\
\frac{1}{(r-p)^2}&=&\sum_{n=0}^\infty N_n(r^2,p^2) U_n(\cos\beta),
\label{denom}\\
A_1(r,q)&=&\sum_{n=0}^\infty a_n(r^2,q^2) U_n(\cos\gamma),\label{A1cheb}\qquad
A_2(r,q)=\sum_{n=0}^\infty b_n(r^2,q^2) U_n(\cos\gamma),\label{A2cheb}
\end{eqnarray}
where 
\begin{eqnarray}
\cos \alpha=\frac{p\cdot q}{p q},\quad
\cos \beta=\frac{p\cdot r}{p r},\quad
\cos \gamma=\frac{q\cdot r}{q r}.
\end{eqnarray}
The vertex functions and kernels $A_1$ and $A_2$
were defined in Eq.~(\ref{vertfiesdef}), respectively, Eq.~(\ref{A1A2def}).
The coefficients $N_n$, $a_n$, and $b_n$ are 
\begin{eqnarray}
N_n(r^2,p^2)=\frac{\theta(r^2-p^2)}{r^2}\left(\frac{p}{r}\right)^n
+\frac{\theta(p^2-r^2)}{p^2}\left(\frac{r}{p}\right)^n,\label{Nn}
\end{eqnarray}
and
\begin{eqnarray}
a_0(r^2,q^2)&=&\frac{1}{2}\left[\left(2-\frac{q^2}{r^2}\right)\theta(r^2-q^2)
+\frac{r^2}{q^2} \theta(q^2-r^2)\right],\label{A0}\\
a_n(r^2,q^2)&=&(-1)^n\frac{(r^2-q^2)}{2}
\left[
\frac{\theta(r^2-q^2)}{r^2}\left( \frac{q}{r}\right)^n
+\frac{\theta(q^2-r^2)}{q^2}\left(\frac{r}{q}\right)^n\right],
\label{an}\qquad n\geq 1,
\end{eqnarray}
and 
\begin{eqnarray}
b_0(r^2,q^2)&=&\frac{1}{2}
\left[\frac{q^2(q^2-3 r^2)}{r^2}\theta(r^2-q^2)
+\frac{r^2(r^2-3q^2)}{q^2} \theta(q^2-r^2)\right],\label{B0}\\
b_1(r^2,q^2)&=&-\frac{1}{2}
\left[
\frac{q^3(q^2-2r^2)}{r^3} \theta(r^2-q^2)
+\frac{r^3(r^2-2q^2)}{q^3} \theta(q^2-r^2)\right],\\
b_n(r^2,q^2)&=&(-1)^n\frac{(r^2-q^2)^2}{2}\left[
\frac{\theta(r^2-q^2)}{r^2}\left( \frac{q}{r}\right)^n 
+\frac{\theta(q^2-r^2)}{q^2}\left(\frac{r}{q}\right)^n\right],\qquad n\geq 2.
\end{eqnarray}

The equations for the scalar vertex functions (\ref{sde_vertfies}) 
and scalar vacuum polarization (\ref{sde_vacpol2}) are expressed
in terms of an infinite set of equations between the harmonics.
Hence
\begin{eqnarray}
\Pi_{\rm S}(q^2)=\frac{1}{4\pi^2}\int\limits_0^{\Lambda^2}
{\rm d} k^2\,\sum_{n=0}^\infty 
\left[ a_n(k^2,q^2) f_n(k^2,q^2)+b_n(k^2,q^2) g_n(k^2,q^2)\right],
\label{vacpolcheb}
\end{eqnarray}
and with Eqs.~(\ref{k11}) and (\ref{k12}) using 
Eqs.~(\ref{denom}) and (\ref{A2cheb}), we get for the harmonics of $F_1$
\begin{eqnarray}
f_l(p^2,q^2)&=&\delta_{0,l}+\frac{\lambda}{(l+1)}
\int\limits_0^{\Lambda^2} {\rm d}r^2\,N_l(r^2,p^2)\nonumber\\
&\times&\sum_{m=0}^\infty
\sum_{n=0}^\infty C_{lmn} \left[ 
a_m(r^2,q^2) f_n(r^2,q^2)+b_m(r^2,q^2) g_n(r^2,q^2)\right], 
\label{coefFl1}
\end{eqnarray}
where
\begin{eqnarray}
C_{lmn}\equiv\frac{2}{\pi}\int\limits_0^\pi{\rm d} \gamma\,
\sin^2\gamma\,U_l(\cos\gamma)U_m(\cos\gamma)U_n(\cos\gamma).
\end{eqnarray}
In the derivation of Eq.~(\ref{coefFl1}) use has been made of 
the fact that
\begin{eqnarray}
\frac{1}{\pi\sin\gamma}\int\limits_{-1}^1{\rm d}\cos\alpha\,
\int\limits_{\cos(\alpha+\gamma)}^{\cos(\alpha-\gamma)}
{\rm d}\cos\beta\, 
U_m(\cos\alpha)U_n(\cos\beta)=
\delta_{m,n}\frac{U_n(\cos\gamma)}{n+1}.
\end{eqnarray}
The symmetric index $C_{lmn}$ can be calculated using product 
properties of the Chebyshev polynomials, giving
\begin{eqnarray}
C_{(2l) mn}&=&\sum_{k=0}^l\left(\delta_{2k,|m-n|}-\delta_{2k,n+m+2}\right),\\
C_{(2l+1) mn}&=&
\sum_{k=0}^l\left(\delta_{2k+1,|m-n|}-\delta_{2k+1,n+m+2}\right),
\qquad l=0,1,2,\dots.
\end{eqnarray}
Then the first two equations for the coefficients $f_n(p^2,q^2)$ 
of vertex function $F_1$ 
read
\begin{eqnarray}
f_0(s,t)&=&1+\lambda
\int\limits_0^{\Lambda^2} {\rm d}u\,N_0(u,s)
\sum_{m=0}^\infty \left[ a_m(u,t) f_m(u,t)+b_m(u,t) g_m(u,t)\right],
\label{f0eq}\\
f_1(s,t)&=&\frac{\lambda}{2}
\int\limits_0^{\Lambda^2} {\rm d}u\,N_1(u,s)\sum_{m=0}^\infty
\biggr[ a_{m+1}(u,t) f_m(u,t)+a_m(u,t) f_{m+1}(u,t)\nonumber\\
&+& b_{m+1}(u,t) g_m(u,t)+b_m(u,t) g_{m+1}(u,t)
\biggr],
\end{eqnarray}
where we have introduced the variables
\begin{eqnarray}
s=p^2,\qquad t=q^2,\qquad u=r^2.
\end{eqnarray}
In principle there is an equivalent set of equations for the 
coefficients $g_n$ of $F_2$.
However we did not succeed in finding explicit expression for these, 
due to our inability to compute explicitly the angular integrals given
by kernels $K_{21}$ and $K_{22}$ of Eqs.~(\ref{k21}) and(\ref{k22}).
The problem is to compute the integrals
\begin{eqnarray}
\int\frac{{\rm d}\Omega_p}{2\pi^2}
\int\frac{{\rm d}\Omega_r}{2\pi^2}  U_n(\cos\alpha) K_{21}(p,q,r),\quad
\int\frac{{\rm d}\Omega_p}{2\pi^2}
\int\frac{{\rm d}\Omega_q}{2\pi^2} U_n(\cos\alpha) K_{21}(p,q,r).
\end{eqnarray}

The main approximation used in this paper is
Eq.~(\ref{canonic}), the replacement of the Yukawa vertex by the zeroth-order
harmonic of $F_1$.
In what follows we estimate the error made by such an approximation.
We define the error $E(q^2)$ in computation of the scalar vacuum 
polarization Eq.~(\ref{vacpolcheb}) as follows:
\begin{eqnarray}
\Pi_{\rm S}(q^2)&=&\frac{1}{4\pi^2}\int\limits_0^{\Lambda^2}
{\rm d} k^2\, a_0(k^2,q^2) f_0(k^2,q^2)+E(q^2),\\
E(q^2)&\equiv&
\frac{1}{4\pi^2}\int\limits_0^{\Lambda^2}
{\rm d} k^2\,\left[
\sum_{n=1}^\infty a_n(k^2,q^2) f_n(k^2,q^2)
+ \sum_{n=0}^\infty b_n(k^2,q^2) g_n(k^2,q^2)\right].\label{error}
\end{eqnarray}
For an estimation of $E(q^2)$ we need to know more about the harmonics
$f_n$, $n\geq 1$, and $g_n$, $n\geq 0$.
The solution to these harmonics is assumed 
to be governed by the harmonic $f_0$ only, 
since the integral over the harmonic $f_0$ acts as the largest inhomogeneous 
term in the integral equations for the higher order harmonics. 
So the equations for the higher order harmonics are approximated by
\begin{eqnarray}
g_n(p^2,q^2)
&\approx&
\lambda \int\limits_0^{\Lambda^2}{\rm d}r^2\,
\int\frac{{\rm d}\Omega_p}{2\pi^2}
\int\frac{{\rm d}\Omega_r}{2\pi^2} 
U_n(\cos\alpha)K_{21}(p,q,r) f_0(r^2,q^2),\label{gneq}\\
f_n(p^2,q^2)&\approx&\frac{\lambda}{(n+1)}
\int\limits_0^{\Lambda^2} {\rm d}r^2\,
N_n(r^2,p^2) a_n(r^2,q^2) f_0(r^2,q^2),\qquad n\geq 1.\label{fneq}
\end{eqnarray}
Unfortunately there is no explicit expression for Eq.~(\ref{gneq}) for the reason 
described above. However it is possible to approximate the angular average by 
considering either one of the three momenta in $K_{21}(p,q,r)$
to be much smaller than the other two. Then the dependence on one of the
three angles between the momenta is lost, and the integration can be performed
explicitly. 
The result for the lowest harmonic of $F_2$, i.e., $g_0$ given by 
Eq.~(\ref{gneq}), is
\begin{eqnarray}
g_0(s,t)
&\approx_{_{_{_{\hspace{-5mm}{{ (s<t)}}}}}}&
\frac{\lambda}{12}
\int_0^s {\rm d}u\,\frac{u}{s^2 t}
 F_{\rm IR}(u,t)+\frac{\lambda}{12}
\int_s^t {\rm d}u\,\frac{1}{s t}
 F_{\rm IR}(u,t)+\frac{\lambda}{12}
\int_t^{\Lambda^2} {\rm d}u\,\frac{1}{u^2}
 F_{\rm UV}(u,t),\\
g_0(s,t)&\approx_{_{_{_{\hspace{-5mm}{{ (s>t)}}}}}}&
\frac{\lambda}{12}
\int_0^t {\rm d}u\,\frac{u}{s^2 t}
 F_{\rm IR}(u,t)+
\frac{\lambda}{12}
\int_t^s {\rm d}u\,\frac{1}{s^2}
 F_{\rm UV}(u,t)+\frac{\lambda}{12}
\int_s^{\Lambda^2} {\rm d}u\,\frac{1}{u^2}
 F_{\rm UV}(u,t),
\end{eqnarray}
and for Eq.~(\ref{fneq}) with Eqs~(\ref{Nn}) and (\ref{an})
\begin{eqnarray}
f_1(s,t)
&\approx_{_{_{_{\hspace{-5mm}{{ (s<t)}}}}}}&
-\frac{\lambda}{4}
\int_0^s {\rm d}u\,
\frac{u-t}{st}\sqrt{\frac{u^2}{st}} 
F_{\rm IR}(u,t)-
\frac{\lambda}{4}\int_s^t {\rm d}u\,
\frac{u-t}{ut}\sqrt{\frac{s}{t}} 
 F_{\rm IR}(u,t)\nonumber\\
&-&\frac{\lambda}{4}
\int_t^{\Lambda^2} {\rm d}u\,
\frac{u-t}{u^2}\sqrt{\frac{st}{u^2}} 
F_{\rm UV}(u,t),\\
f_1(s,t)&\approx_{_{_{_{\hspace{-5mm}{{ (s>t)}}}}}}&
-\frac{\lambda}{4}
\int_0^t {\rm d}u\,\frac{u-t}{st}\sqrt{\frac{u^2}{st}} 
 F_{\rm IR}(u,t)-\frac{\lambda}{4}
\int_t^s {\rm d}u\,
\frac{u-t}{us}
\sqrt{\frac{t}{s}} 
F_{\rm UV}(u,t)\nonumber\\
&-&\frac{\lambda}{4}
\int_s^{\Lambda^2} {\rm d}u\,
\frac{u-t}{u^2}\sqrt{\frac{st}{u^2}} 
 F_{\rm UV}(u,t),
\end{eqnarray}
where $s=p^2$, $t=q^2$, and we have used Eq.~(\ref{channelapprox}).
These equations can be analyzed in detail once the solutions
for the channel functions, Eqs.~(\ref{fir2}) and (\ref{fuv2}), are known.
But for obtaining the asymptotic behavior of the harmonics $g_0$ and $f_1$
it is sufficient to use the asymptotics of the channels, i.e., take
$F_{\rm IR}(p^2,q^2)\rightarrow 
F_{\rm IR}(0,q^2)\propto (q^2/\Lambda^2)^{\omega/2-1/2}$,
$F_{\rm UV}(p^2,q^2)\rightarrow 
F_{\rm UV}(p^2,0)\propto (p^2/\Lambda^2)^{\omega/2-1/2}$.
This gives for the harmonics 
\begin{eqnarray}
g_0(p^2,q^2)&\propto&\lambda \frac{1}{p^2} F_{\rm UV}(p^2,0),\qquad
f_1(p^2,q^2)\propto\lambda \frac{q}{p} F_{\rm UV}(p^2,0),
\qquad p^2\gg q^2,\label{fguv_est}\\
g_0(p^2,q^2)&\propto& \lambda\frac{1}{p^2} F_{\rm IR}(0,q^2),
\qquad
f_1(p^2,q^2)\propto\lambda \frac{p}{q} F_{\rm IR}(0,q^2),
\qquad q^2\gg p^2.\label{fgir_est}
\end{eqnarray}
The above equations give the leading behavior 
(in either $q/p \ll 1$ or $p/q\ll 1$) of the harmonics 
$g_0$, $f_1$ in terms of $f_0$ up to some $\lambda$-dependent 
factor, which is ${\cal O}(1)$ (thus nonsingular in $\lambda$).
Furthermore, from Eq.~(\ref{fneq}) we get the relation
\begin{eqnarray}
\frac{f_{n+1}(p^2,q^2)}{f_n(p^2,q^2)}
\propto-\frac{q}{p},\qquad p^2\gg q^2,\qquad
\frac{f_{n+1}(p^2,q^2)}{f_n(p^2,q^2)}
\propto
-\frac{p}{q},\qquad q^2\gg p^2,
\end{eqnarray}
and we assume a similar relation to hold between the harmonics 
$g_{n+1}$ and $g_n$.
Thus the series  
\begin{eqnarray}
S(p^2,q^2)
\equiv\sum_{m=0}^\infty \left[ a_m(p^2,q^2) f_m(p^2,q^2)
+b_m(p^2,q^2) g_m(p^2,q^2)\right],\label{S}
\end{eqnarray}
which occurs both in the equation for $f_0$, Eq.~(\ref{f0eq}), 
and for $\Pi_{\rm S}$, Eq.~(\ref{vacpolcheb}), will be rapidly converging for 
either $p^2\gg q^2$ or $p^2\ll q^2$, since 
\begin{eqnarray}
\frac{a_{n+1}(p^2,q^2)f_{n+1}(p^2,q^2)}{a_n(p^2,q^2)f_n(p^2,q^2)}
\approx  \frac{\min(p^2,q^2)}{\max(p^2,q^2)}\ll 1,
\end{eqnarray}
and again a similar equation for the part containing the harmonics $g_n$.
At $p^2=q^2$ only three terms of the series $S$, Eq.~(\ref{S}), 
contribute, since $a_n(p^2,p^2)=0$ for $n\geq 1$ 
and $b_n(p^2,p^2)=0$ for $n\geq 2$.
Hence, a straightforward approximation for the series $S$ is
\begin{eqnarray}
S(p^2,q^2)&\approx&a_0(p^2,q^2) 
f_0(p^2,q^2)
+{\cal O}\left(a_1(p^2,q^2) f_1(p^2,q^2)\right)
+{\cal O}\left(b_0(p^2,q^2) g_0(p^2,q^2)\right),
\end{eqnarray}
supporting Eq.~(\ref{canonic}).
With the expressions 
obtained for $f_1$ and $g_0$, Eqs.~(\ref{fguv_est}) and (\ref{fgir_est}), 
the leading term of the error $E$ defined 
in Eq.~(\ref{error}) can be estimated.
The leading term of the error $E(q^2)$ is given by
\begin{eqnarray}
E(q^2)&\approx&\frac{1}{4\pi^2}\int\limits_0^{\Lambda^2}{\rm d}k^2\,
\left[a_1(k^2,q^2)f_1(k^2,q^2)+b_0(k^2,q^2)g_0(k^2,q^2)\right]\nonumber\\
&\sim& \lambda\int\limits_{q^2}^{\Lambda^2}{\rm d}k^2\,
\frac{q^2}{k^2} F_{\rm UV}(k^2,q^2=0)
+\mbox{next-to-leading}
\nonumber\\
&\sim&
\Lambda^2\left[
(1+{\cal O}(\lambda))\left(\frac{q^2}{\Lambda^2}\right)^{\omega/2+1/2}
-(1+{\cal O}(\lambda))\frac{q^2}{\Lambda^2} \right],\label{error_expr}
\end{eqnarray}
where we have kept only leading terms
and for $F_{\rm UV}(p^2,0)=F_1(p,p)$ given by Eq.~(\ref{GSqzero}).
Recall that $\omega=\sqrt{1-4\lambda}$.
The estimation of Eq.~(\ref{error_expr}) can be checked more explicitly 
by using the solutions obtained in Sec.~\ref{sec_scal_vertex}
for the channel functions $F_{\rm IR}$, $F_{\rm UV}$, 
Eqs.~(\ref{fir2}) and (\ref{fuv2}).

Eq.~(\ref{error_expr}) shows that when $\lambda=0$, $\omega=1$, 
the error $E$ vanishes, and when
$\omega <1$ clearly the terms in the error can be neglected with respect to
the first two terms on the right-hand side of
$\Pi_{\rm S}(q^2)$, see Eq.~(\ref{vacasym2}).
Thus this analysis supports the assumption Eq.~(\ref{canonic})
made in Sec.~\ref{sec_scal_vertex}
and the error $E$ contributes only to next-to-next-to-leading
order in $q^2/\Lambda^2$.
And therefore we may conclude that our 
approximation gives correct leading and next-to-leading 
behavior of $\Pi_{\rm S}(q^2)$.

\end{document}